\documentclass[12pt]{article}
\usepackage{amsmath,graphicx,enumerate,natbib,url,setspace,lscape,longtable}
\usepackage{epsfig,mathrsfs,amsthm,amssymb,color, verbatim,multirow,makecell} 
\usepackage{authblk,float,array,lscape,bm,comment,setspace}
\usepackage{amsfonts,indentfirst,booktabs,caption,subfig,blindtext,comment,placeins,mathabx}
\usepackage[ruled,vlined]{algorithm2e}  
\usepackage[colorlinks=true]{hyperref}
\hypersetup{
	linkcolor= blue, 
	citecolor= blue 
}

\def\1{\mathbf{1}}

\def\mR{\mathbb{R}}

\def\hat{\widehat}

\def \tbe_t{\tilde{\bm{\epsilon}}_t}

\newtheorem{theorem}{Theorem}
\newtheorem{lemma}{Lemma}
\newtheorem{remark}{Remark}
\newtheorem{proposition}{Proposition}
\newtheorem{corollary}{Corollary}

\def\beq{\begin{equation}}
	\def\eeq{\end{equation}}
\def\ben{\begin{equation*}}
	\def\een{\end{equation*}}
\def\bea{\begin{eqnarray}}
	\def\eea{\end{eqnarray}}
\def\bda{\begin{eqnarray*}}
	\def\eda{\end{eqnarray*}}

\def\bet{\begin{theorem}}
	\def\eet{\end{theorem}}
\def\bel{\begin{lemma}}
	\def\eel{\end{lemma}}
\def\bep{\begin{proposition}}
	\def\eep{\end{proposition}}
\def\bec{\begin{corollary}}
	\def\eec{\end{corollary}}
\def\bc{\begin{center}}
	\def\ec{\end{center}}

\graphicspath{{Fig/}}

\newcommand{\blind}{1}

\numberwithin{equation}{section}

\begin{document}

\def\spacingset#1{\renewcommand{\baselinestretch}%
	{#1}\small\normalsize} \spacingset{1}


\if1\blind
{
	\title{\bf A spectral inference method for determining the number of communities in networks}
	\author[1]{Yujia Wu\thanks{Email: \texttt{yujwu@polyu.edu.hk}}}
	\author[2]{Xiucai Ding\thanks{Email: \texttt{xcading@ucdavis.edu.}  XCD is partially supported by NSF DMS-2515104.}}
	\author[3]{Jingfei Zhang\thanks{Email: \texttt{emma.zhang@emory.edu}}}
	\author[4]{Wei Lan\thanks{Email: \texttt{lanwei@swufe.edu.cn}}}
	\author[5]{Chih-Ling Tsai\thanks{Email: \texttt{cltsai@ucdavis.edu}}}
	\affil[1]{Department of Data Science and Artificial Intelligence, The Hong Kong Polytechnic University}
	\affil[2]{Department of Statistics, University of California, Davis}
	\affil[3]{Goizueta Business School, Emory University}
	\affil[4]{Laboratory for Data Intelligence and Innovation in Finance and Economics and Center of Statistical Research, Southwestern University of Finance and Economics}
	\affil[5]{Graduate School of Management, University of California, Davis}
	\renewcommand*{\Affilfont}{\small}  
	\renewcommand*{\Authands}{ and }  
	\date{}  
	\maketitle
} \fi

\if0\blind
{
	\bigskip
	\bigskip
	\bigskip
	\begin{center}
		{\LARGE\bf A spectral inference method for determining the number of communities in networks}
	\end{center}
	\medskip
} \fi
\bigskip

\begin{abstract}
	
To characterize the community structure in network data, researchers have developed various block-type models, including the stochastic block model, the degree-corrected stochastic block model, the mixed membership block model, the degree-corrected mixed membership block model, and others.
A critical step in applying these models effectively is determining the number of communities in the network. However, to the best of our knowledge, existing methods for estimating the number of network communities either rely on explicit model fitting or fail to simultaneously accommodate network sparsity and a diverging number of communities. In this paper, we propose a model-free spectral inference method based on eigengap ratios that addresses these challenges. The inference procedure is straightforward to compute, requires no parameter tuning, and can be applied to a wide range of block models without the need to estimate network distribution parameters. Furthermore, it is effective for both dense and sparse networks with a divergent number of communities. Technically, we show that the proposed spectral test statistic converges to a {function of the type-I Tracy-Widom distribution via the Airy kernel} under the null hypothesis, and that the test is asymptotically powerful under weak alternatives. Simulation studies on both dense and sparse networks demonstrate the efficacy of the proposed method. Three real-world examples are presented to illustrate the usefulness of the proposed test.

\end{abstract}

\noindent%
{\it Keywords:} Network data, community detection, spectral inference,  eigengap-ratio test, Tracy-Widom distribution

	\spacingset{1.3} 

\section{Introduction} \label{sec-introduction}
\noindent  {Since the identification of community structures in networks by \cite{lorrain1971structural} and \cite{granovetter1973strength}, various block models have been introduced to characterize the block structures underlying network generation mechanisms and to enhance the effective utilization of network data. To name but a few,  
	the stochastic block model (SBM) \citep{holland1983stochastic,abbe2018community}, the degree-corrected stochastic block model (DCSBM) \citep{karrer2011stochastic,jin2023optimal}, the mixed membership model (MM) \citep{airoldi2008mixed,ma2021determining} and the degree-corrected mixed membership model (DCMM) \citep{jin2017estimating,han2023universal,jin2024mixed}. 
	
	In these models, the network data are modeled as undirected random graphs without self-loops. They assume that the associated adjacency matrix $A = (A_{ij}) \in \mR^{n \times n}$ is symmetric, with all diagonal elements equal to zero. In addition,  the network is assumed to be generated from a Bernoulli-distributed random graph model; that is,  
	\beq\label{Ber}
	A_{ij} \sim \text{Bernoulli}(P_{ij}), \quad \text{for every } i < j,
	\eeq
	where $P = (P_{ij}) \in \mR^{n \times n}$ is the symmetric edge-probability matrix. Specifically, $P_{ij}$ denotes the probability of a connection between nodes $i$ and $j$, for $i \neq j$. In the aforementioned block models—SBM, DCSBM, MM, and DCMM—the main differences lie in the specific assumptions and formulations imposed on $P$. Nevertheless, a common and essential feature is that the probability matrix $P$ typically has rank {$\mathsf{K}$} \citep{jin2023optimal}, where {$\mathsf{K}$} also identifies the number of communities.
	
	In practice, however, the number of communities {$\mathsf{K}$} is often unknown {\it a priori} and must be estimated before analyzing specific block models. In this paper, we aim to develop a theoretically sound and practically efficient sequential testing approach that is universally applicable in the sense that it does not depend on the structural and regularity assumptions of any specific block model on $P$ and remains valid even when {$\mathsf{K}$} diverges with $n$. In what follows, we first summarize the existing approaches in Section \ref{sec_summaryliterature}, and then provide an overview of our proposed method in Section \ref{sec_overview}. }

\subsection{Summary of existing literature}\label{sec_summaryliterature}


\noindent  
As discussed above, the existing literature on estimating {$\mathsf{K}$} largely relies on fitting specific block models. To streamline our review and presentation, we focus on the DCMM, which encompasses the other popular network block models mentioned above. The DCMM can be formulated as follows. 

{Let $Q = (Q_{kl}) \in \mathbb{R}^{\mathsf{K} \times \mathsf{K}}$ be a symmetric community interaction matrix. Let $\boldsymbol{\omega} = (\omega_1, \ldots, \omega_n)^\top$ denote the vector of node-specific degree parameters that capture degree heterogeneity, where $\omega_i > 0$ for all $i$ and $\sum_{i=1}^n \omega_i = n$.}
In addition, each node $i$ is associated with a ${\mathsf{K}}$-dimensional  probability mass vector
\[
\bm{\pi}_i=(\pi_{i,1},\cdots,\pi_{i,{\mathsf{K}}})^\top,\quad \text{with}\ \pi_{i,k}\ge0 \ \text{and}\ \sum_{k=1}^{\mathsf{K}} \pi_{i,k}=1,
\]
where $\pi_{i,k}$ represents the probability that node $i$ belongs to community $k$.
The DCMM assumes that for $P_{ij}$ in (\ref{Ber}) follows 
\beq\label{def-DCMM}
P_{ij}=\omega_i\omega_j\,\bm{\pi}_i^\top Q \bm{\pi}_j,\quad \text{for}\ 1\le i\le j\le n.
\eeq

Note that the SBM, DCSBM, and MM are all special cases of the above unified formulation. Specifically, the DCMM reduces to the SBM when $\omega_i\equiv 1$ and each $\bm{\pi}_i$ is degenerate, i.e., $\pi_{i,k}\in \{0,1\}$; it reduces to the DCSBM when each $\bm{\pi}_i$ is degenerate while $\omega_i$ varies across nodes; and it reduces to the MM when $\omega_i\equiv 1$ and $\pi_{i,k}\in(0,1)$ allows overlapping community memberships.

In recent years, a variety of inference-based methods have been developed to estimate ${\mathsf{K}}$ under the aforementioned block models, subject to various regularity conditions. The literature can be broadly divided into two classes: sequential testing methods (c.f. (\ref{test-hypo}))  and non-sequential testing methods. For the sequential testing methods, \citet{lei2016goodness} develops a goodness-of-fit test for the SBM based on the largest eigenvalue of a standardized adjacency matrix, while \citet{hu2021using} proposes entry-wise deviation statistics for goodness-of-fit tests under both the SBM and DCSBM. However, both studies focus on the dense regime, in the sense that they assume the edge probabilities $P_{ij}$ are of constant order  for all $1 \le i,j \le n$, in order to ensure certain universality in the standardized adjacency matrix appearing in their test statistics.  In the sparse regime when {$\max_{i, j}P_{ij} \downarrow 0$,}  \citet{jin2023optimal} proposes a stepwise goodness-of-fit (StGoF) procedure for optimally estimating the ${\mathsf{K}}$ in SBM and DCSBM when ${\mathsf{K}}$ is assumed to be fixed. We note that all of the above methods require prior estimation of the unknown network parameters $\bm{\pi}_i$, $Q$, and $\bm{\omega}$ in order to compute their test statistics before determining ${\mathsf{K}}$. For possibly divergent ${\mathsf{K}}$, when ${\mathsf{K}=\mathrm{O}(\log\log n)}$, \citet{han2023universal} develops a rank inference via residual subsampling procedure for testing ${\mathsf{K}}$ that is  applicable under \eqref{def-DCMM}. However, the RIRS procedure involves a tuning parameter that must be carefully chosen to ensure efficient subsampling when constructing the test statistic. For the readers' convenience, a summary of these limitations for sequential-based inference is provided in Table~\ref{Comparison}.

%

\begin{table}[!ht]
	\setlength{\belowcaptionskip}{0.2cm}
	\scriptsize
	\centering
	\resizebox{1.0\textwidth}{!}{
		\begin{tabular}{cccccc}
			\hline {Method}
			&\multicolumn{1}{c}{\makecell{Dense networks \\with $P_{ij} \asymp 1$}}&  \multicolumn{1}{c}{\makecell{Sparse networks \\
					with $P_{ij} \ll 1$}} & \multicolumn{1}{c}{\makecell{Diverging number \\
					of communities}}& \multicolumn{1}{l}{\makecell{No parameter\\
					estimation (model-free)}} & \multicolumn{1}{l}{\makecell{No tuning \\parameters}}\\
			\cline{1-6}
			\hline
			Proposed method &$\checkmark$ &$\checkmark$ &$\checkmark$ &$\checkmark$ &$\checkmark$ \\
			\cite{lei2016goodness} &$\checkmark$ &$\times$ &$\checkmark$ &$\times$ &$\checkmark$\\
			\cite{hu2021using} &$\checkmark$ &$\times$ &$\checkmark$ &$\times$ &$\checkmark$\\
			\cite{han2023universal} &$\checkmark$ &$\checkmark$ &{$\checkmark$} &$\checkmark$ &$\times$\\
			\cite{jin2023optimal} &$\checkmark$ &$\checkmark$ &$\times$ &$\times$ &$\checkmark$\\
			\bottomrule
	\end{tabular}} \caption{Comparisons of our proposed method below and other sequential inference methods. The symbol ``$\checkmark$" means applicable, and ``$\times$" stands for not applicable. }
	\label{Comparison}
\end{table}

We note that there also exist non-sequential testing methods, including network cross-validation approaches \citep{chen2018network} and likelihood-based methods \citep{daudin2008mixture,wang2017likelihood,saldana2017many,noroozi2019estimation,hu2020corrected,ma2021determining}. Notably, in addition to the significant computational complexity, these methods exhibit limitations similar to those summarized in Table~\ref{Comparison}, particularly for likelihood-based approaches \citep{wang2017likelihood,noroozi2019estimation,hu2020corrected,ma2021determining,jin2023optimal}.

Motivated by the above challenges, in this paper we propose a sequential spectral inference–based method that does not rely on specific model assumptions, allows the network to be either dense or sparse, and permits ${\mathsf{K}}$ to diverge with $n$. An overview of the proposed method is provided in Section~\ref{sec_overview}.


\subsection{An overview of our proposed method}\label{sec_overview}


\noindent  
In this paper, we develop a spectral inference procedure for determining ${\mathsf{K}}$ based on a sequential testing framework by considering the following hypothesis:
\beq\label{test-hypo}
\mathbf{H}_{0}: {\mathsf{K}=\mathsf{K}_0} \quad \text{v.s.}\quad {\mathbf{H}_{1}: {\mathsf{K}_0 < \mathsf{K} \leq \mathsf{K}_{\max}},  }
\eeq
where ${\mathsf{K}}$ denotes the unknown true rank of $P$, ${\mathsf{K}_0}$ is the hypothesized rank of $P$, and ${\mathsf{K}_{\max}}$ is a possible upper bound on ${\mathsf{K}}$. Here we consider a one-sided alternative, since a network with ${\mathsf{K}>\mathsf{K}_0}$ communities can also be represented as a network with ${\mathsf{K}_0>\mathsf{K}}$ communities by splitting one or more true communities. This one-sided alternative has also been considered in \cite{lei2016goodness}, \cite{chen2018network}, \cite{wang2017likelihood}, and \cite{hu2021using}. Note that through (\ref{test-hypo}), a natural consistent sequential testing estimate of ${\mathsf{K}}$ can be generated, that is, 
\begin{equation}\label{eq_estimator}
	{\widehat{\mathsf{K}}}:=\inf \{ {\mathsf{K}_0} \geq 0: \ \mathbf{H}_0 \ \text{is} \ \text{accepted}\}. 
\end{equation}

To address the challenges discussed in Section~\ref{sec_summaryliterature}, we present in Algorithm~\ref{alg:ER-test2} an inference procedure for~\eqref{test-hypo} based on selected eigenvalues of the adjacency matrix~$A$. The proposed method does not rely on any specific model assumptions on~$P$ and allows ${\mathsf{K}}$ to diverge. It is theoretically sound, computationally efficient, and does not require the involvement of tuning parameters. To test~\eqref{test-hypo}, for a suitably chosen $\mathsf{K}_{\max}$, our proposed algorithm utilizes the following test statistic
\beq\label{test-T}
{\mathbb{T}=\frac{\lambda_{\mathsf{K}_0+1}(A)-\lambda_{\mathsf{K}_{\max}+1}(A)}{\lambda_{\mathsf{K}_{\max}+1}(A)-\lambda_{\mathsf{K}_{\max+2}}(A)}, }
\eeq
where $\lambda_1(A) \ge \lambda_2(A) \ge \cdots \ge \lambda_n(A)$ denote the eigenvalues of $A$ arranged in decreasing order. As will be justified by our theoretical results in Section~\ref{sec-property}, the distribution of ${\mathbb{T}}$ under $\mathbf{H}_0$ is characterized by some functional form of the type-I Tracy–Widom distribution via the Airy kernel. Under $\mathbf{H}_1$, ${\mathbb{T}}$ diverges at the rate {$\mathrm{O}(n^{2/3})$,} which makes it a powerful test statistic for (\ref{test-hypo}).

 In practice, constructing $\mathbb{T}$ requires specifying an upper bound $\mathsf{K}_{\max}$. Theoretically, as established in Theorem \ref{theory-null}, one may set $\mathsf{K}_{\max} = \mathsf{K}_0 + \mathsf{C}$ for a sufficiently large fixed integer $\mathsf{C}$. As demonstrated in our simulations and real data analyses, the proposed method is robust to the choice of $\mathsf{K}_{\max}$, provided it is selected within a reasonable range. For automated implementation, Section \ref{sec-method} presents a practical, data-driven procedure for choosing $\mathsf{K}_{\max}$ based on the permutation method discussed in \cite{Dobriban2020}.

We point out that similar, though quite different, eigen-ratio statistics have been used for estimating the number of factors or signals via large sample covariance matrices when $\mathsf{K}$ is assumed to be fixed; see \cite{onatski2009formal,bao2015universality,ding2022tracy}. We refer readers to Remark~\ref{rmk_eigenratio} for a more detailed discussion on the difference.

Although the limiting distribution of ${\mathbb{T}}$ can be established theoretically, the unknown and potentially complicated structure of $P$ prevents it from admitting an explicit and compact formula for deriving critical values under $\mathbf{H}_0$. To address this issue, we further show that the target distribution can be accurately mimicked via calibration by constructing suitable quantities from carefully selected eigenvalues of an $n \times n$ symmetric Wigner matrix with mean zero and Gaussian entries of variance $1/n$, that is, a Gaussian Orthogonal Ensemble (GOE) matrix. Consequently, the critical values can be accurately and efficiently determined via calibration. Unlike many existing methods, we emphasize that our proposed algorithm does not involve the procedure of tuning parameters, and is therefore computationally efficient. As will be demonstrated in Sections~\ref{sec-simu} and~\ref{sec-realdata}, our method outperforms existing approaches in both simulation studies and real data analyses in terms of accuracy, power, and computational efficiency.

As shown in Table \ref{Comparison}, our proposed method accommodates both dense and sparse networks and allows ${\mathsf{K}}$ to diverge with $n$. Generally speaking, denser networks permit a larger number of communities. To rigorously characterize this trade-off and establish the asymptotic distribution of ${\mathbb{T}}$, we impose a sufficient condition that involves both the sparsity level and the number of communities. More explicitly, we require that $n^{1/3} \max_{i, j} P_{ij}/{\mathsf{K}^2} \gg 1$ as an explicit trade-off; see Figure \ref{fig:tradeoff} of the supplement for an illustration. This condition recovers the upper bound ${\mathsf{K}} \ll n^{1/6}$ for dense SBMs in \cite{lei2016goodness}. Moreover, such a condition can be easily satisfied under~\eqref{def-DCMM} for commonly used block models. Deriving optimal conditions and developing results and methods for extremely sparse networks are beyond the scope of the current paper and will be explored in future works.

The remainder of this article is organized as follows. Section~\ref{sec-method} introduces the proposed algorithm, and Section~\ref{sec-property} establishes its theoretical properties. Section~\ref{sec-simu} presents simulation studies, with results supporting the advantages of the proposed method summarized in Table~\ref{Comparison}.  {Section~\ref{sec-realdata} reports two empirical analyses, and Section~\ref{sec-conclude} provides concluding remarks.}  All technical details and additional discussions and numerical results are deferred to the supplementary file.

\section{Our proposed algorithm}\label{sec-method}

\noindent
In this section, we formally introduce our proposed algorithm. As discussed in Section~\ref{sec_overview}, the algorithm relies on the construction of the test statistic $\mathbb{T}$ in \eqref{test-T}, which requires specifying an upper bound $\mathsf{K}_{\max}$. Theoretically (cf.\ Theorem~\ref{theory-null}), given the hypothesized testing rank $\mathsf{K}_0$, the validity of our procedure is guaranteed as long as $\mathsf{K}_{\max}-\mathsf{K}_0$ remains bounded. For practical implementation, we adopt the permutation-based approach (also known as parallel analysis), as studied in \cite{Dobriban2020}, which has been widely used for rank determination in factor analysis based on sample covariance matrices and is computationally feasible and efficient.

The procedure is summarized below, with full details provided in Algorithm~\ref{alg:PA} in the supplementary material for the reader’s convenience. Briefly, given the adjacency matrix $A$, one constructs permuted matrices {$A_{\bm\tau}$} by independently shuffling the entries within each column of $A$ and repeating this permutation procedure multiple times. An eigenvalue of $A$, is selected if it exceeds a specified percentile (e.g., $95\%$) of the corresponding eigenvalues obtained from the permuted matrices {$A_{\bm\tau}$}.  Let $\mathsf{K}_{\text{PA}}$ denote the number of eigenvalues selected by the parallel analysis. According to the discussion in \cite{Dobriban2020}, with high probability one has $\mathsf{K}_{\text{PA}} \ge \mathsf{K}$. Motivated by this result, to obtain a practical upper bound, we consistently set $\mathsf{K}_{\max} = \mathsf{K}_{\text{PA}} + 5$ in our simulations and real data analyses. This automated choice performs well in practice.

With the above preparation, we now present our proposed method in Algorithm \ref{alg:ER-test2} for testing \eqref{test-hypo}. In Step 1, we construct the test statistic \eqref{test-T} using $\mathsf{K}_{\max}$ selected via Algorithm \ref{alg:PA}, as discussed above. In Step 2, motivated by the theoretical results in Theorem \ref{theory-null}, we compute the critical value using $\mathsf{J}$ independent Gaussian Orthogonal Ensemble (GOE) matrices. For convenience, we choose the significance level $\alpha$ and $\mathsf{J}$ such that $(1-\alpha)\mathsf{J}$ is an integer.  

\begin{algorithm}[!ht]

\caption{The proposed spectral inference procedure}
\label{alg:ER-test2}

\KwIn{Adjacency matrix $A$, ${\mathsf{K}_0}$, the number of synthetic GOE matrices ${\mathsf{J}}$ (e.g., $\mathsf{J} = 1000$), and the significance level $\alpha$.}
\noindent {\bf 1. Construction of the test statistic.}
\begin{enumerate}
	\item[(i).] Calculate $\mathsf{K}_{\max}$ from Algorithm \ref{alg:PA}.
	\item[(ii).] Construct the test statistic $\mathbb{T}$ in (\ref{test-T}). 
\end{enumerate}	
\noindent {\bf 2. Calibration of the critical value.} 
\begin{enumerate}	
	
	\item[(i).]	\For{$\mathsf{j} = 1$ \KwTo ${\mathsf{J}}$}{
		\begin{enumerate}
			\item[(a).]	 Generate an GOE matrix $W_{\mathsf{j}}$ that $W_{\mathsf{j}}=W_{\mathsf{j}}^\top$,  where the off-diagonal entries $W_{\mathsf{j}, k<l}\stackrel{\text{i.i.d.}}{\sim}\mathcal{N}(0, n^{-1})$ and the diagonal entries $W_{\mathsf{j}, kk}\stackrel{\text{i.i.d.}}{\sim}\mathcal{N}(0, 2/n)$ . 		
			
			\item[(b).]	Calculate the calibration statistic
			\begin{equation}\label{eq_calibrationstatistic}
				\mathbb{T} _{W_{\mathsf{j}}} = \frac{\lambda_1(W_{\mathsf{j}}) - \lambda_{\mathsf{K}_{\max}-\mathsf{K}_0+1}(W_{\mathsf{j}})}{\lambda_{\mathsf{K}_{\max}-\mathsf{K}_0+1}(W_{\mathsf{j}}) - \lambda_{\mathsf{K}_{\max}-\mathsf{K}_0+2}(W_{\mathsf{j}})}.
			\end{equation}
	\end{enumerate} }
	
	\item[(ii).]	Compute the critical value $c_\alpha$ as the $(1-\alpha)$ quantile of the ${\mathsf{J}}$ simulated values ${\{\mathbb{T}_{W_{\mathsf{1}}}, \cdots, \mathbb{T}_{W_{\mathsf{J}}}\}}$.
\end{enumerate}	

\KwOut{Reject the null hypothesis $\mathbf{H}_0$ in (\ref{test-hypo}) if $\mathbb{T} > c_\alpha$.}

\end{algorithm}

%
%

In what follows, we provide heuristic discussion and remarks on the validity of the proposed algorithm. The rigorous theoretical analysis is deferred to Section \ref{sec-property}. Suppose the true rank is $\mathsf{K}$. Under $\mathbf{H}_0$ in (\ref{test-hypo}), we show that the eigengaps appearing in both the numerator and the denominator are of the same order, implying that the statistic $\mathbb{T}$ in (\ref{test-T}) is of order $\mathrm{O}(1)$. In contrast, under the alternative hypothesis $\mathsf{K} > \mathsf{K}_0$, the numerator becomes significantly larger than the denominator, and consequently $\mathbb{T}$ grows on the order of $\mathrm{O}(n^{2/3})$. This separation renders the test statistic powerful. To select an appropriate critical value and ensure accurate testing, we establish the asymptotic distribution of $\mathbb{T}$, which does not admit a closed-form expression. Nevertheless, this distribution can be accurately mimicked by a statistic generated from Gaussian Orthogonal Ensemble (GOE) matrices, provided that the eigenvalues are properly chosen, as described in (\ref{eq_calibrationstatistic}).

We note that our method is also computationally feasible. On the one hand, Step 2—calculating the critical values—can be carried out offline, with the resulting values tabulated for use in the testing procedure. On the other hand, the method requires computing only the largest $\mathsf{K}_{\max}+2$ eigenvalues of $A$, for which many efficient numerical linear algebra algorithms are available, enabling fast computation, especially for sparse networks. For further discussion of computational aspects and a comparison of computational time across different methods, we refer the reader to Section \ref{sec_computationaleffiency} of the supplement.

\begin{remark}\label{rmk_eigenratio}
We note that, although eigen-ratio–based test statistics have been explored in inference and estimation problems for large sample covariance matrices—see, for example, \citep{onatski2009formal, ding2021spiked, ding2021spikedAos, bao2015universality, ding2022tracy, ding2025tracy}—these studies typically assume that $\mathsf{K}$ is fixed in (\ref{test-hypo}). In contrast, our proposed statistic and testing procedure, developed for network adjacency matrices, are very different.

First, in the aforementioned works, the ratio is often constructed using gaps between the consecutive eigenvalues, whereas our statistic considers eigenvalue gaps with larger index separations. This design allows $\mathsf{K}$ to diverge in general. Second, many existing approaches require the eigenvalues of the population matrix to be distinct due to the specific construction of their statistics, while our method imposes no such restriction on the matrix $P$. Finally, although the asymptotic distributions in these works can also be mimicked using synthetic matrices, they typically rely on rectangular Wishart matrices and require selecting an aspect ratio, which serves as a tuning parameter. In contrast, our calibration is based on Gaussian Orthogonal Ensemble (GOE) matrices and does not involve such tuning. 
\end{remark}

\section{Theoretical justification}\label{sec-property}

\noindent
In this section, we establish the theoretical properties of the proposed method. In Section \ref{sec_asymptoticdistribution}, we derive the asymptotic distribution of the proposed test statistic $\mathbb{T}$ in \eqref{test-T} under the null hypothesis $\mathbf{H}_0$. Section \ref{sec_power} then investigates the power of the test and justifies the performance of the associated estimator in \eqref{eq_estimator}. Throughout the paper, we will consistently use the following conventions. For sequences $\{a_n\}$ and $\{b_n\}$ depending on $n$, $a_n=\mathrm{O}(b_n)$ means $|a_n|\le \mathsf{C}|b_n|$ for some constant $\mathsf{C}>0$, while $a_n=\mathrm{o}(b_n)$  means $|a_n|\le c_n|b_n|$ for some positive sequence $c_n\downarrow0$ as $n\to\infty$. We also use  $a_n\asymp b_n$ to mean both $a_n=\mathrm{O}(b_n)$ and $b_n=\mathrm{O}(a_n)$.

\subsection{Asymptotic distribution of $\mathbb{T}$}\label{sec_asymptoticdistribution}

\noindent
We first introduce some well-known results in random matrix theory. Consider a Gaussian Orthogonal Ensemble (GOE) matrix $W \in \mathbb{R}^{n \times n}$, where $W_{ij} = W_{ji}$, the off-diagonal entries $W_{ij}$ for $i \neq j$ are independent and distributed as $\mathcal{N}(0, n^{-1})$, and the diagonal entries $W_{ii}$ are distributed as $\mathcal{N}(0, 2/n)$. It is well known that, after appropriate normalization, the largest eigenvalue of $W$, namely $n^{2/3}(\lambda_1(W)-2)$, converges in distribution to the type-I Tracy–Widom distribution introduced in \cite{tracy1994level,tracy1996orthogonal}. Moreover, for any fixed $\mathsf{C} > 0$, the joint distribution of the first $\mathsf{C}$ largest eigenvalues of $W$, after normalization $\{ n^{2/3}(\lambda_k(W) - 2) \}_{1 \leq k \leq \mathsf{C}}$, admits a joint density described by the Airy kernel, also commonly referred to as the Tracy–Widom law; see, for example, \cite{forrester1993spectrum}. 

With the above preparation, we now state the asymptotic distribution of $\mathbb{T}$ with the aid of the GOE matrix $W$. Throughout the remainder of this section, we adopt the standard assumption in network analysis of \citep{bickel2009nonparametric,zhao2012consistency} that {$\max_{i,j} P_{ij} \asymp \min_{i, j} P_{ij}$.} The main result of this section is stated below.

\noindent
%


\bet\label{theory-null}
Assume that  {$n^{1/3}\max_{i, j}P_{ij}/\mathsf{K}^2\to\infty$ as $n\to\infty$, and $\min_{1\le k\le \mathsf{K}}|\lambda_{k}(P)|\ge cn\max_{i, j}P_{ij}/\mathsf{K}$}  for some constant $c>0$. 
Then under the null hypothesis $\mathbf{H}_0$ in (\ref{test-hypo}), when $\mathsf{K}_{\max}-\mathsf{K}_0$ is finite, we have
$\lim_{n \rightarrow \infty} F_{\mathbb{T}}(x)=\lim_{n \rightarrow \infty} F_{\mathbb{T}_W}(x)$ for any $x\in\mR$,
where $F_{\mathbb{T}}(x)$ is the cumulative distribution function for the test statistic $\mathbb{T}$, and $F_{\mathbb{T}_W}(x)$ is the cumulative distribution function of $\mathbb{T}_W$, defined as
$$
\mathbb{T}_W=\frac{\lambda_1(W)-\lambda_{\mathsf{K}_{\max}-\mathsf{K}_0+1}(W)}{\lambda_{\mathsf{K}_{\max}-\mathsf{K}_0+1}(W)-\lambda_{\mathsf{K}_{\max}-\mathsf{K}_0+2}(W)},
$$
where $W$ is an GOE matrix.
\eet

On the one hand, Theorem \ref{theory-null} shows that, under certain conditions, the distribution of the test statistic $\mathbb{T}$ can be asymptotically characterized by that of $\mathbb{T}_W$. Note that since $\mathbb{T}_W$ can be rewritten as follows
$$
\mathbb{T}_W=\frac{n^{2/3}(\lambda_1(W)-2)-n^{2/3}(\lambda_{\mathsf{K}_{\max}-\mathsf{K}_0+1}(W)-2)}{n^{2/3}(\lambda_{\mathsf{K}_{\max}-\mathsf{K}_0+1}(W)-2)-n^{2/3}(\lambda_{\mathsf{K}_{\max}-\mathsf{K}_0+2}(W)-2)},
$$
then according to the discussion at the beginning of this section, the quantity of interest is essentially a function of the type-I Tracy–Widom distribution, which is characterized by the Airy kernel. On the other hand, this representation also enables us to compute the critical value for testing \eqref{test-hypo} via calibration. Specifically, for a given nominal level $\alpha$, let $c_\alpha = F_{\mathbb{T}_W}^{-1}(1-\alpha)$ denote the upper $\alpha$-quantile of $\mathbb{T}_W$. The critical value $c_\alpha$ is then obtained by numerically approximating $F_{\mathbb{T}_W}$, for example through Monte Carlo simulation as in Algorithm \ref{alg:ER-test2}, and we reject $\mathbf{H}_0$ whenever $\mathbb{T} > c_\alpha$. Several remarks regarding the assumptions of the theorem are in order.

\begin{remark}  We point out that our condition 
{$n^{1/3}\max_{i, j}P_{ij} / \mathsf{K}^2 \to \infty$} establishes an explicit trade-off between network sparsity, measured by {$\max_{i, j} P_{ij}$,} and the possible divergent value of $\mathsf{K}$; see Figure \ref{fig:tradeoff} of the supplement for an illustration. On the one hand, this condition allows {$\max_{i, j} P_{ij} \downarrow 0$} when $\mathsf{K}$ is fixed or grows slowly, thereby accommodating sparse networks. As a result, it is less restrictive than the conditions in \citet{lei2016goodness} and \citet{hu2021using}, which assume dense networks with $P_{ij} \asymp 1$ for all $1 \le i,j \le n$. On the other hand, when the network is denser, the same condition permits $\mathsf{K} \to \infty$. In particular, it recovers the regime $\mathsf{K} \ll n^{1/6}$ for dense stochastic block models considered in \citet{lei2016goodness}, and is less restrictive than the fixed $\mathsf{K}$ assumption in \citet{jin2023optimal} and the very slowly diverging $\mathsf{K}$ (e.g., $\mathrm{O}(\log\log n)$) considered in \citet{han2023universal}.

Nevertheless, this condition is primarily technical and is imposed to ensure that, under $\mathbf{H}_0$ in (\ref{test-hypo}), the distribution of the statistic $\mathbb{T}$ can be described by the Tracy–Widom law; see Remark \ref{sec_techinicaldiscussion} of our supplement for more details.  When this condition fails—either in extremely sparse networks, as considered in \citet{han2023universal} and \citet{jin2023optimal}, or when the number of communities grows sufficiently fast, as in \citet{hu2021using}—our current method is likely to break down. The underlying technical reason is analogous to that discussed in \citet{huang2020transition}: the proposed statistic may no longer follow a Tracy–Widom distribution, but instead converge to a Gaussian limit or to a free additive convolution of Gaussian and Tracy–Widom components. Since a rigorous treatment of these regimes lies beyond the scope of the present paper, we leave their investigation to future work.

\end{remark}

\begin{remark}
We note that, in the current paper, we also impose an assumption on the smallest nonzero eigenvalues of $P$. Heuristically, this assumption can be interpreted as a signal-to-noise ratio condition, ensuring that the spectral behavior (and hence the rank) of $P$ can be reliably inferred from the leading eigenvalues of the adjacency matrix $A$. Similar conditions have been adopted in various settings; see, for example, \citep{lei2015consistency,jin2023optimal}. We also refer readers to Remark~\ref{rmk_techinicalremark} of  our supplement for further details. Such an assumption is easily satisfied by certain block models under mild conditions. For instance, in the well-studied assortative stochastic block model, one has $\min_{1 \le k \le \mathsf{K}} |\lambda_k(P)| \asymp {n(a-b)}/\mathsf{K}$ \citep{abbe2018community,Amini2018}, where $a$ and $b$ denote the within- and between-community connection probabilities. In this setting, our assumption holds automatically. Similar arguments apply to the degree-corrected stochastic block model when the node-specific degree parameters are of constant order.
%
%
%
%
%
%

\end{remark}

\subsection{Analysis of the power and the sequential estimator}\label{sec_power} 

\noindent
In this section, we investigate the power of the proposed statistic $\mathbb{T}$ and, as a consequence, provide an asymptotic analysis of the estimator (\ref{eq_estimator}). As discussed earlier, under $\mathbf{H}_1$ in (\ref{test-hypo}), the conditions of Theorem~\ref{theory-null} no longer hold, and $\mathbb{T}$ is expected to take large values, thereby exhibiting strong discriminatory power against the alternative hypothesis. The result is stated as follows.


\bet\label{theory-alter}
Assume that the conditions of Theorem~\ref{theory-null} hold and that $\mathbf{H}_1$ in (\ref{test-hypo}) is true. Then, for any arbitrarily large constant $\mathsf{C}>0$, we have
\begin{equation*}
\mathbb{P}(\mathbb{T}> \mathsf{C} n^{2/3})=1-\mathrm{o}(1).
\end{equation*}
\eet

This theorem implies that, for a given nominal level $\alpha$, one has $\mathbb{P}(\mathbb{T} > c_\alpha) = 1 - \mathrm{o}(1)$ under the alternative hypothesis $\mathbf{H}_1$, and hence the null hypothesis $\mathbf{H}_0$ will be rejected with probability tending to one. In particular, Theorem~\ref{theory-alter} indicates that $\mathbb{T}$ diverges at least at the rate $\mathrm{O}(n^{2/3})$ under the alternative, making our test more powerful than those proposed in \cite{han2023universal} and \cite{hu2021using}. Specifically, under the alternative, the test statistic in \cite{han2023universal} remains bounded away from zero, while the test statistic in \cite{hu2021using} diverges only at the order $\mathrm{O}(\log n)$. Moreover, the test statistic of \cite{hu2021using} is known to lack power under the planted partition stochastic block model, namely, a stochastic block model with equal community sizes, equal $Q_{ll}$’s, and equal $Q_{lk}$’s for $l \neq k$.

With the help of Theorems~\ref{theory-null} and \ref{theory-alter}, we can analyze the associated estimator $\widehat{\mathsf{K}}$ generated according to (\ref{eq_estimator}). Under our testing framework, this is equivalent to defining
$$\widehat{\mathsf{K}} = \min \big\{ \mathsf{K}_0 \in \{1,\ldots,\mathsf{K}_{\max}\} : \mathbb{T} \le c_\alpha \big\}.$$
The properties of this estimator are summarized as follows.


\bec \label{cor:hatK-consistency}
Under the assumptions of Theorems~\ref{theory-null} and \ref{theory-alter}, we have that  
\[
\mathbb{P}\left( \hat{\mathsf{K}}= \mathsf{K} \right)= 1-\alpha-\mathrm{o}(1). 
\]

\eec

This corollary establishes that $\widehat{\mathsf{K}}$ effectively estimates the true number of communities $\mathsf{K}$ at the nominal level $\alpha$. In particular, as $\alpha \downarrow 0$ (equivalently, as $c_\alpha=\mathrm{o}(n^{2/3})$ diverges), Corollary~\ref{cor:hatK-consistency} implies that $\mathbb{P}\left(\widehat{\mathsf{K}} = \mathsf{K}\right)=1-\mathrm{o}(1)$, thereby yielding a consistent estimator for $\mathsf{K}$. However, in practical sequential testing procedures, the nominal level $\alpha$ is typically kept fixed rather than allowed to vanish with $n$, as letting $\alpha \downarrow 0$ often results in unreliable empirical size performance numerically. Motivated by this consideration, we can follow \cite{lei2016goodness,ding2022tracy} to consider the estimator
$$\widehat{\mathsf{K}}^* = \min \big\{ \mathsf{K}_0 \in \{1,\ldots,\mathsf{K}_{\max}\} : \mathbb{T} \le t_n \big\},$$
where $t_n \to \infty$ as $n \to \infty$ and is independent of the nominal level $\alpha$. The properties of $\widehat{\mathsf{K}}^*$ are established in the following corollary.

\bec \label{cor:Kstar-consistency}

Under the assumptions of Theorems~\ref{theory-null} and \ref{theory-alter}, when $t_n$ satisfies that $t_n \asymp n^{\epsilon}$ for some $\epsilon \in (0,5/6)$, we have that
\[
\mathbb{P}\left( \hat{\mathsf{K}}^*= \mathsf{K} \right)=1-\mathrm{o}(1).
\]

\eec


\section{Numerical simulations}\label{sec-simu}
\noindent
To evaluate the finite-sample performance of Algorithm \ref{alg:ER-test2}, we conduct simulation studies based on \eqref{def-DCMM} to test the hypotheses in \eqref{test-hypo} under both dense and sparse network settings under various setups.
 
To generate $\bm{\pi}_i$, we follow \cite{han2023universal} and simulate them from one of the two sets: $\mathbf{PM}_{\mathsf{K}}=\{\bm{e}_1,\cdots,\bm{e}_\mathsf{K}\},$ where $\bm{e}_k$ denotes the $k$-th standard basis vector in $\mathbb{R}^{\mathsf{K}}$, and $\mathbf{MM}_{\mathsf{K}}=\Big\{(0.2,0.8,\underbrace{0,\cdots,0}_{\mathsf{K}-2}),(0.8,0.2,\underbrace{0,\cdots,0}_{\mathsf{K}-2})$,
		$(\underbrace{\mathsf{K}^{-1},\cdots,\mathsf{K}^{-1}}_\mathsf{K})\Big\}$. To generate $\omega_i$, we consider two settings:  $\omega_i\equiv 1$ for all $1 \leq i \leq n$, or following  \cite{zhao2012consistency}, $\omega_i's$ are independently drawn from a distribution with unit expectation:
\begin{equation}\label{eq_omegasettingtwo}
\omega_i=\left\{
\begin{array}{lr}
\eta_i,& \text{w.p.}\quad 0.8,  \\
9/11,& \text{w.p.}\quad 0.1, \\
13/11,& \text{w.p.}\quad 0.1,
\end{array}
\right.
\end{equation}
where $\eta_i$ is uniformly distributed on the interval $[4/5,6/5]$. The setups for the $Q$ matrices are given in Sections \ref{subsec-densesimu} and \ref{subsec-spasesimu}, which characterize the dense and sparse network settings, respectively.

{
With the above setups, to generate networks according to (\ref{def-DCMM}) with block structures, we consider three different variants of commonly used block models.
\begin{enumerate}
\item Stochastic block model (SBM). We set $\bm\pi_i\in \mathbf{PM}_{\mathsf{K}}$ and  $\omega_i\equiv1.$ For each community $k$, we randomly select $n/\mathsf{K}$ nodes and set their corresponding  $\bm\pi_i=\bm{e}_k$, so that $\sum_{i=1}^n\pi_{i,k}=n/\mathsf{K}$.
\item Degree-corrected stochastic block model (DCSBM). We set $\bm\pi_i\in \mathbf{PM}_{\mathsf{K}}$ with $\omega_i$ generated according to (\ref{eq_omegasettingtwo}). For each community $k$, we randomly select $n/\mathsf{K}$ nodes and set their corresponding  $\bm\pi_i=\bm{e}_k$, so that $\sum_{i=1}^n \pi_{i,k}=n/\mathsf{K}$.
\item  Degree-corrected mixed membership model (DCMM). We set  $\bm \pi_i\in \mathbf{PM}_{\mathsf{K}}\cup \mathbf{MM}_{\mathsf{K}}$ with  $\omega_i$ generated according to (\ref{eq_omegasettingtwo}). For each community $k$, following \cite{han2023universal}, we randomly select $n_0:=(\mathsf{K}^{-1}-0.03)n$ nodes and assign their $\bm\pi_i$ to $\bm{e}_k$. This results in a total of $n_0\mathsf{K}$ nodes whose $\bm\pi_i$ are drawn from $\mathbf{PM}_{\mathsf{K}}$. For the remaining nodes, we randomly allocate $(n-n_0\mathsf{K})/3$ nodes for each vector in  $\mathbf{MM}_{\mathsf{K}}$ and assign their $\bm\pi_i$ accordingly. 
\end{enumerate} 


In all simulations, the network size is set to $n=3,000$ with $\mathsf{K}\in\{3, 5, 10, 15, 20\}$. The dense networks are studied in Section \ref{subsec-densesimu}, while the sparse networks are analyzed in Section \ref{subsec-spasesimu}. 
} We compare our proposed statistic and method in Algorithm \ref{alg:ER-test2} (denoted as $\mathbb{T}$) with five other methods: the two methods in \cite{lei2016goodness},  denoted as  $\mathbb{T}_{\text{Lei}}$ and  $\mathbb{T}_{\text{Lei,boot}}$, the method in  \cite{han2023universal}, denoted as $\mathbb{T}_{\text{Han}}$, and the two methods in  \cite{hu2021using}, denoted as $\mathbb{T}_{\text{Hu}}$ and $\mathbb{T}^{\text{aug}}_{\text{Hu,boot}}$, respectively. Here, $\mathbb{T}_{\text{Lei,boot}}$ and $\mathbb{T}^{\text{aug}}_{\text{Hu,boot}}$ denote tests obtained by applying a bootstrap correction and an augmented bootstrap procedure to $\mathbb{T}_{\text{Lei}}$ and $\mathbb{T}_{\text{Hu}}$, respectively. Since the methods of \cite{lei2016goodness} and \cite{hu2021using} are not applicable for DCMM, we omit them in simulation studies for DCMM. 


\subsection{Performance for dense networks}\label{subsec-densesimu}

\noindent
In this section, we compare the performance of various methods on dense networks, following the setups in \cite{hu2021using,han2023universal} for generating $Q$. For dense SBM and DCSBM, we set $Q_{kl}=0.1(1+4\times \mathbb{I}(k=l))$, while for dense DCMM we take $Q_{kl}=0.1^{|k-l|}$ if $k\neq l$, and $Q_{kl}=(\mathsf{K}+1-k)/\mathsf{K}$ otherwise.

\noindent

Tables~\ref{Tab:dense2-sbmQ1}--\ref{Tab:dense2-dcmmQ2} report the empirical sizes and powers for dense SBM, dense DCSBM  and dense DCMM. Overall, the simulation studies demonstrate that our proposed test performs well across all three block models under dense networks and outperforms other methods. These findings are consistent with our theoretical results. We elaborate on these findings in more detail below.

For the results of dense SBM and DCSBM presented in Tables~\ref{Tab:dense2-sbmQ1} and \ref{Tab:dense2-dcsbmQ1}, the empirical sizes of {$\mathbb{T}$, $\mathbb{T}_{\text{Han}}$ and $\mathbb{T}^{\text{aug}}_{\text{Hu, boot}}$} are close to the true 5\% nominal level,
although {$\mathbb{T}^{\text{aug}}_{\text{Hu, boot}}$ performs slightly worse than $\mathbb{T}$ and $\mathbb{T}_{\text{Han}}$. For other methods, under SBM, when $\mathsf{K}>5$,} the sizes of {$\mathbb{T}_{\text{Hu}}$, $\mathbb{T}_{\text{Lei}}$ and $\mathbb{T}_{\text{Lei, boot}}$} deviate notably from the nominal level. Under DCSBM, when {$\mathsf{K}>5$, $\mathbb{T}_{\text{Hu}}$} exhibits size distortions, which can be much larger than 5\%. Moreover, for both SBM and DCSBM, the empirical sizes of {$\mathbb{T}_{\text{Lei}}$ and $\mathbb{T}_{\text{Lei,boot}}$} deviate significantly from the 5\% nominal level, except for smaller $\mathsf{K}=3$. 
Second, the empirical powers of {$\mathbb{T}$ and $\mathbb{T}^{\text{aug}}_{\text{Hu, boot}}$ are generally equal to 1, with $\mathbb{T}$ performing more powerfully than $\mathbb{T}^{\text{aug}}_{\text{Hu, boot}}$, while $\mathbb{T}_{\text{Hu}}$ is less powerful. Notably, $\mathbb{T}_{\text{Han}}$} shows almost no power.
In addition, under SBM, the powers of {$\mathbb{T}_{\text{Lei}}$, and $\mathbb{T}_{\text{Lei, boot}}$ are generally equal to 1. However, under DCSBM, due to the size distortions of $\mathbb{T}_{\text{Lei}}$ and $\mathbb{T}_{\text{Lei, boot}}$}, we do not further analyze their empirical powers.

For the dense DCMM results reported in Table~\ref{Tab:dense2-dcmmQ2}, on the one hand, the empirical sizes of $\mathbb{T}$ are close to the nominal 5\% level. In contrast, $\mathbb{T}_{\text{Han}}$ exhibits noticeable size distortion when $\mathsf{K}>5$, indicating that the test $\mathbb{T}_{\text{Han}}$ may not be suitable for large values of $\mathsf{K}$.
On the other hand, the empirical powers of $\mathbb{T}$ increase as $\mathsf{K}-\mathsf{K}_0$ grows. Since the sizes of $\mathbb{T}_{\text{Han}}$ are distorted and its empirical powers are unstable for $\mathsf{K}>5$, we do not further analyze its empirical power performance.

\begin{table}[!ht]
\centering
\renewcommand\arraystretch{1.5}{
	\scriptsize
	\begin{tabular}{cccccc|ccccc}
		\hline
		& \multicolumn{5}{c}{$\mathbb{T}$} &\multicolumn{5}{c}{$\mathbb{T}_{\text{Han}}$} \\
		\cline{1-11}
		$\mathsf{K}$ &3 & 5& 10& 15 & 20 &3 & 5& 10& 15 & 20\\
		\hline
		$\mathsf{K}_0=3$ &0.052&1&1&1&1    &0.052&0.068&0.044&0.038&0.026\\
		$\mathsf{K}_0=5$ &*&0.060&1&1&1      &*&0.044&0.030&0.048&0.044\\
		$\mathsf{K}_0=10$ &* &* &0.042&1&1          &* &* &0.048&0.030&0.046\\
		$\mathsf{K}_0=15$ &*&* &*&0.058&1    &* &* &* &0.038&0.040\\
		$\mathsf{K}_0=20$ &*&* &* &* &0.048    &*  &*  &*  &* &0.046 \\
		\hline
		&\multicolumn{5}{c}{$\mathbb{T}_{\text{Hu}}$}& \multicolumn{5}{c}{$\mathbb{T}^{\text{aug}}_{\text{Hu,boot}}$}\\
		\hline
		$\mathsf{K}$ &3 & 5& 10& 15 & 20 &3 & 5& 10& 15 & 20\\
		\hline
		$\mathsf{K}_0=3$ &0.068&0.110&0.828&0.994&1  &0.066&0.994&1&1&1\\
		$\mathsf{K}_0=5$ &*&0.080&0.580&0.976&1        &*&0.042&1&1 &1    \\
		$\mathsf{K}_0=10$ &* &* &0.130&0.500&0.986      &* &* &0.066&1&1  \\
		$\mathsf{K}_0=15$ &*  &*   &* &0.242&0.560    &*&* &* &0.034&0.914 \\
		$\mathsf{K}_0=20$  &*    &* &*   &* &0.398      &*  &*  &*  &* &0.044 \\
		\hline
		&\multicolumn{5}{c}{$\mathbb{T}_{\text{Lei}}$}& \multicolumn{5}{c}{$\mathbb{T}_{\text{Lei,boot}}$}\\
		\cline{1-11}
		$\mathsf{K}$ &3 & 5& 10& 15 & 20 &3 & 5& 10& 15 & 20\\
		\hline
		$\mathsf{K}_0=3$ &0.064&1&1&1&1   &0.052&1&1&1&1 \\
		$\mathsf{K}_0=5$ &*&0.072&1&1 &1          &*&0.096&1&1&1\\
		$\mathsf{K}_0=10$ &* &* &0.182&1 &1     &* &* &0.138&1&1\\
		$\mathsf{K}_0=15$ &*  &* &* &0.614&1        & *  &*  &* &0.730&1    \\
		$\mathsf{K}_0=20$ &*    &* &*  &* &0.846     &*  &*   &*   &* &0.930  \\
		\bottomrule
\end{tabular}}
\caption{Empirical sizes and powers for dense SBM under the nominal level $\alpha = 0.05.$}
\label{Tab:dense2-sbmQ1}
\end{table}

\begin{table}[!ht]
\centering
\renewcommand\arraystretch{1.5}{
	\scriptsize
	\begin{tabular}{cccccc|ccccc}
		\hline
		& \multicolumn{5}{c}{$\mathbb{T}$} &\multicolumn{5}{c}{$\mathbb{T}_{\text{Han}}$} \\
		\cline{1-11}
		$\mathsf{K}$ &3 & 5& 10& 15 & 20 &3 & 5& 10& 15 & 20\\
		\hline
		$\mathsf{K}_0=3$ &0.036&1&1&1 &1   &0.046&0.054 &0.044  &0.036&0.050\\
		$\mathsf{K}_0=5$ &*&0.056&1&1&1      &*&0.064&0.050&0.052&0.046\\
		$\mathsf{K}_0=10$ &*&*&0.048&1&1         &*&*  &0.054&0.050&0.050\\
		$\mathsf{K}_0=15$ &*&*&*& 0.034 &1    &* &  *  &* & 0.050 &0.054 \\
		$\mathsf{K}_0=20$ &*&*&*&* &0.044    &*   &*   &*   &* &0.052 \\
		\hline
		&\multicolumn{5}{c}{$\mathbb{T}_{\text{Hu}}$}& \multicolumn{5}{c}{$\mathbb{T}^{\text{aug}}_{\text{Hu,boot}}$}\\
		\hline
		$\mathsf{K}$ &3 & 5& 10& 15 & 20 &3 & 5& 10& 15 & 20\\
		\hline
		$\mathsf{K}_0=3$ &0.036&0.276 &0.840&0.988&1  &0.028&0.854&1 &1&1\\
		$\mathsf{K}_0=5$ &*&0.042&0.466&0.842&1          &*&0.052&0.890&1&1   \\
		$\mathsf{K}_0=10$ &*&*  &0.182 &0.766&1    &*&* &0.072 &0.900&1  \\
		$\mathsf{K}_0=15$ &*   &*    &* &0.304 &1     &*   &*   &* &0.062 &0.990 \\
		$\mathsf{K}_0=20$  &*   &*   &*   &* &0.712    &*   &*   &*   &* &0.040\\
		\hline
		&\multicolumn{5}{c}{$\mathbb{T}_{\text{Lei}}$}& \multicolumn{5}{c}{$\mathbb{T}_{\text{Lei,boot}}$}\\
		\cline{1-11}
		$\mathsf{K}$ &3 & 5& 10& 15 & 20 &3 & 5& 10& 15 & 20\\
		\hline
		$\mathsf{K}_0=3$ &0.110&1&1 &0.994&0.990   &0.064&1&0.588 &0.572&0.646 \\
		$\mathsf{K}_0=5$ &*&1&0.936&0.978&0.984   &*&0.006&0&0.008&0.688  \\
		$\mathsf{K}_0=10$ &*&*  &0.268 &0.820&0.810   &*&*  &0.084 &0.012&0.006 \\
		$\mathsf{K}_0=15$ &* &*   &* &0.114&0.612   &*   &*   &*   &0.014 &0   \\
		$\mathsf{K}_0=20$ &*   &*   &*   &* &0.162   &*   &*   &*   &* &0.026   \\
		\bottomrule
\end{tabular}}
\caption{Empirical sizes and powers for dense DCSBM under the nominal level $\alpha = 0.05.$}
\label{Tab:dense2-dcsbmQ1}
\end{table}

\begin{table}[!ht]
\centering
\renewcommand\arraystretch{1.5}{
	\scriptsize
	\begin{tabular}{cccccc|ccccc}
		\hline
		& \multicolumn{5}{c}{$\mathbb{T}$} &\multicolumn{5}{c}{$\mathbb{T}_{\text{Han}}$} \\
		\hline
		$\mathsf{K}$ &3 & 5& 10& 15 & 20 &3 & 5& 10& 15 & 20\\
		\hline
		$\mathsf{K}_0=3$ &0.058&0.344& 0.610&0.864 &0.966  &0.034 &1&1&0.420&0.802\\
		$\mathsf{K}_0=5$ &* &0.056 & 0.330& 0.440& 0.646& *&0.040& 1&0.408&0.768\\
		$\mathsf{K}_0=10$ &* &* &0.038 &0.184 &0.390 &* & * &0.112&0.414&0.766\\
		$\mathsf{K}_0=15$  & * & * & * &0.040 &0.186  &  *& * & * &0.342&0.740 \\
		$\mathsf{K}_0=20$ & * & * & *& *&0.036  & * & * & *& *&0.688\\
		\bottomrule
\end{tabular}}
\caption{Empirical sizes and powers for dense DCMM under the nominal level $\alpha = 0.05.$}
\label{Tab:dense2-dcmmQ2}
\end{table}

\subsection{Performance for sparse networks}\label{subsec-spasesimu}

\noindent
In this section, we compare the performance of various methods on sparse networks. Analogous to Section~\ref{subsec-densesimu}, for sparse SBM and DCSBM, we set  $Q_{kl}=0.1n^{-2/9}(1+4\times \mathbb{I}(k=l))$, while for sparse DCMM we take $Q_{kl}=0.5n^{-2/9}0.1^{|k-l|}$ if $k\not = l$, and $Q_{kl}=0.5n^{-2/9}(\mathsf{K}+1-k)/\mathsf{K}$ otherwise.

Tables~\ref{Tab:sparse2-sbmQ1}--\ref{Tab:sparse2-dcmmQ2} report the empirical sizes and powers for sparse SBM, sparse DCSBM  and sparse DCMM. Overall, the simulation studies demonstrate that our proposed method performs well across the three block models under sparse networks and outperforms other methods.  We provide a more detailed discussion of these results below.

For the sparse SBM and DCSBM results reported in Tables~\ref{Tab:sparse2-sbmQ1} and \ref{Tab:sparse2-dcsbmQ1}, the empirical sizes of $\mathbb{T}$ and $\mathbb{T}_{\text{Han}}$ are close to the true nominal 5\% level. In contrast, although $\mathbb{T}_{\text{Lei,boot}}$ and $\mathbb{T}^{\text{aug}}_{\text{Hu,boot}}$ achieve reasonable empirical sizes in a few settings, their performance is unstable, and most of their empirical sizes exhibit noticeable distortions. Moreover, $\mathbb{T}_{\text{Hu}}$ and $\mathbb{T}_{\text{Lei}}$ deviate substantially from the nominal level. Turning to power, the empirical power of $\mathbb{T}$ increases steadily as $\mathsf{K}-\mathsf{K}_0$ grows. By contrast, $\mathbb{T}_{\text{Han}}$ exhibits almost no power. Since the empirical sizes of $\mathbb{T}_{\text{Lei}}$, $\mathbb{T}_{\text{Lei,boot}}$, $\mathbb{T}_{\text{Hu}}$, and $\mathbb{T}^{\text{aug}}_{\text{Hu,boot}}$ are largely distorted, we do not further analyze their empirical powers.

For the sparse DCMM results reported in Table~\ref{Tab:sparse2-dcmmQ2}, on the one hand, the empirical sizes of $\mathbb{T}$ are close to the nominal 5\% level, whereas the empirical sizes of $\mathbb{T}_{\text{Han}}$ become distorted when $\mathsf{K} > 5$. On the other hand, the empirical powers of $\mathbb{T}$ increase as $\mathsf{K}-\mathsf{K}_0$ grows. Since the sizes of $\mathbb{T}_{\text{Han}}$ are distorted and its empirical powers are unstable for $\mathsf{K} > 5$, we refrain from further analyzing the powers of $\mathbb{T}_{\text{Han}}$.

%

\begin{table}[!ht]
\centering
\renewcommand\arraystretch{1.5}{
\scriptsize
\begin{tabular}{cccccc|ccccc}
	\hline
	& \multicolumn{5}{c}{$\mathbb{T}$} &\multicolumn{5}{c}{$\mathbb{T}_{\text{Han}}$} \\
	\hline
	$\mathsf{K}$ &3 & 5& 10& 15 & 20 &3 & 5& 10& 15 & 20\\
	\hline
	$\mathsf{K}_0=3$ &0.054&0.634&0.792&0.916&0.968    &0.052&0.064&0.042 &0.038  &0.052\\
	$\mathsf{K}_0=5$ &*&0.048&0.766&0.810&0.948     &*&0.056&0.070&0.068&0.046\\
	$\mathsf{K}_0=10$ &*&*&0.054& 0.520 &0.840         &*&*&0.040&0.046&0.058\\
	$\mathsf{K}_0=15$ &* &*&* &0.052 &0.290    &*  &*  &* &0.044  &0.062\\
	$\mathsf{K}_0=20$ &*&* &* &*  &0.038     &*  &*  &*  &* &0.042 \\
	\hline
	&\multicolumn{5}{c}{$\mathbb{T}_{\text{Hu}}$}& \multicolumn{5}{c}{$\mathbb{T}^{\text{aug}}_{\text{Hu,boot}}$}\\
	\cline{1-11}
	K &3 & 5& 10& 15 & 20 &3 & 5& 10& 15 & 20\\
	\hline
	$\mathsf{K}_0=3$ &0.258&0.830&0.990&1&1  &0.060&0.596&0.992&1&1\\
	$\mathsf{K}_0=5$ &*&0.380&1&1&1                 &*&0.040&0.840&0.972&1 \\
	$\mathsf{K}_0=10$ &*&*&0.992&1&1          &*&*&0.016&0.848 &1  \\
	$\mathsf{K}_0=15$ &* &* &* &1 &1        &*&* &* &0.026 &0.982 \\
	$\mathsf{K}_0=20$  &* &* &* &* &1      &* &* &* &* &0.088\\
	\hline
	&\multicolumn{5}{c}{$\mathbb{T}_{\text{Lei}}$}& \multicolumn{5}{c}{$\mathbb{T}_{\text{Lei,boot}}$}\\
	\hline
	$\mathsf{K}$ &3 & 5& 10& 15 & 20 &3 & 5& 10& 15 & 20\\
	\hline
	$\mathsf{K}_0=3$  &0.990&0.994&1&1&1  &0.080 &0.976&0.850&0.982&1 \\
	$\mathsf{K}_0=5$   &*&1&1&1&1     &*&0.162&0.636&0.840&1 \\
	$\mathsf{K}_0=10$  &*&*&1&1 &1      &*&*&0.016&0.932&1 \\
	$\mathsf{K}_0=15$  &* &* &*  &1 &1     &* &* &* &0.074&1 \\
	$\mathsf{K}_0=20$  &*  &*  &* &* &1     &*  &*  &* &*  &0 \\
	\bottomrule
	\end{tabular}}
\caption{Empirical sizes and powers for sparse SBM under the nominal level $\alpha = 0.05.$}
\label{Tab:sparse2-sbmQ1}
\end{table}

\begin{table}[!ht]
\centering
\renewcommand\arraystretch{1.5}{
\scriptsize
\begin{tabular}{cccccc|ccccc}
	\hline
	& \multicolumn{5}{c}{$\mathbb{T}$} &\multicolumn{5}{c}{$\mathbb{T}_{\text{Han}}$} \\
	\cline{1-11}
	$\mathsf{K}$ &3 & 5& 10& 15 & 20 &3 & 5& 10& 15 & 20\\
	\hline
	$\mathsf{K}_0=3$  &0.040&0.410&0.544&0.890&1 &0.040&0.052&0.058&0.042&0.050\\
	$\mathsf{K}_0=5$  &*&0.054&0.250&0.694&0.900     &*&0.064&0.060&0.056&0.048 \\
	$\mathsf{K}_0=10$  &*&*&0.046 &0.512&0.620        & *& *&0.048&0.038&0.040\\
	$\mathsf{K}_0=15$  &* & *&*&0.042&0.510    &* &*& *&0.066&0.052\\
	$\mathsf{K}_0=20$  &*&*&* &* &0.036     &* & *& *& *& 0.046 \\
	\hline
	&\multicolumn{5}{c}{$\mathbb{T}_{\text{Hu}}$}& \multicolumn{5}{c}{$\mathbb{T}^{\text{aug}}_{\text{Hu,boot}}$}\\
	\cline{1-11}
	$\mathsf{K}$ &3 & 5& 10& 15 & 20 &3 & 5& 10& 15 & 20\\
	\hline
	$\mathsf{K}_0=3$ &0.020&0.830&0.932&0.756&0.650  &0.060&0.484&0.802&0.910&1\\
	$\mathsf{K}_0=5$ &*&0.130&1&1&1  &*&0.168&0.750&0.626&0.942 \\
	$\mathsf{K}_0=10$ & *& *&0.650 &1&1      &* &*&0.058 &0.542&1 \\
	$\mathsf{K}_0=15$ &* &* &*&0.808 &1        &* &* &*&0.250&0.854\\
	$\mathsf{K}_0=20$  &* &* &* &*& 1   &* &* &* &* &0.026\\
	\hline
	&\multicolumn{5}{c}{$\mathbb{T}_{\text{Lei}}$}& \multicolumn{5}{c}{$\mathbb{T}_{\text{Lei,boot}}$}\\
	\cline{1-11}
	$\mathsf{K}$ &3 & 5& 10& 15 & 20 &3 & 5& 10& 15 & 20\\
	\hline
	$\mathsf{K}_0=3$  &0.230&1&1&1 &1  &0.034&0.862&0.644&0.850&0.456 \\
	$\mathsf{K}_0=5$   &*&0.168&1&1&1  &*&0.180&0.590&0.248 &0.832 \\
	$\mathsf{K}_0=10$  &* &*&0.642 &1&1     & * & *&0.076 &0.340&0.250 \\
	$\mathsf{K}_0=15$  &* &* &*&0.136 &1    &* &* &*&0.034&1  \\
	$\mathsf{K}_0=20$  &*  &*  &* &* &0     &*  &*  &* &*  &0  \\
	\bottomrule
	\end{tabular}}
\caption{Empirical sizes and powers for sparse DCSBM under the nominal level $\alpha = 0.05.$}
\label{Tab:sparse2-dcsbmQ1}
\end{table}

\begin{table}[!ht]
\centering
\renewcommand\arraystretch{1.5}{
\scriptsize
\begin{tabular}{cccccc|ccccc}
	\hline
	& \multicolumn{5}{c}{$\mathbb{T}$} &\multicolumn{5}{c}{$\mathbb{T}_{\text{Han}}$} \\
	\hline
	$\mathsf{K}$ &3 & 5& 10& 15 & 20 &3 & 5& 10& 15 & 20\\
	\hline
	$\mathsf{K}_0=3$ &0.044&0.286 & 0.460& 0.638& 0.872 &0.048 &1&1&0.910& 0.956\\
	$\mathsf{K}_0=5$ &* &0.038 & 0.250&0.482 &0.600 & *&0.040& 1&0.900 &0.952\\
	$\mathsf{K}_0=10$ &* &* &0.052 &0.206 & 0.390&* & * &0.410&0.898&0.944\\
	$\mathsf{K}_0=15$  & * & * & * &0.038 & 0.154 &  *& * & * &0.786 &0.938 \\
	$\mathsf{K}_0=20$ & * & * & *& *&0.040  & * & * & *& *&0.900\\
	\bottomrule
	\end{tabular}}
\caption{Empirical sizes and powers for sparse DCMM under the nominal level $\alpha = 0.05.$}
\label{Tab:sparse2-dcmmQ2}
\end{table}

	\section{Real data analyses}\label{sec-realdata}
	\noindent
In this section, we illustrate the practical utility of our proposed method through the analysis of two real-world networks—the political blog network \cite{adamic2005political} in Section~\ref{subsec-blog} and the Sina Weibo network \cite{wu2022inward} in Section~\ref{subsec-sina}—and compare its performance with that of existing methods. An additional application to a Facebook network from Simmons College is presented in Section~\ref{apen-sec5} of the supplementary material.


	\subsection{Political blog network}\label{subsec-blog}
	
	\noindent
	The first dataset records links between internet blogs over the two-month period preceding the 2004 U.S. presidential election and was originally studied by \cite{adamic2005political}. The nodes represent blogs, and the edges correspond to web links between them. Following standard practice in prior studies \citep{lei2016goodness,hu2021using,han2023universal}, we focus on the largest connected component of the network, which consists of 1,222 nodes and has an edge density of approximately 2.24\%.

Since all blogs fall into either “conservative” or “liberal” communities based on political stance, the true number of communities in this network can be regarded as $\mathsf{K}=2$. We apply our test statistic $\mathbb{T}$, along with the competing methods $\mathbb{T}_{\text{Lei, boot}}$, $\mathbb{T}_{\text{Hu, boot}}^{\text{aug}}$, and $\mathbb{T}_{\text{Han}}$, to test $\mathsf{K}_0 = 1$ and $2$. The testing results, summarized in Table~\ref{political}, show that $\mathbb{T}$, $\mathbb{T}_{\text{Han}}$, and $\mathbb{T}_{\text{Hu, boot}}^{\text{aug}}$ reject $\mathbf{H}_0: \mathsf{K} = 1$ but do not reject $\mathbf{H}_0: \mathsf{K} = 2$. In contrast, $\mathbb{T}_{\text{Lei, boot}}$ rejects all null hypotheses. Therefore, the results of $\mathbb{T}$, together with those of $\mathbb{T}_{\text{Han}}$ and $\mathbb{T}_{\text{Hu, boot}}^{\text{aug}}$, are consistent and support the ground truth that $\mathsf{K}=2$.

\begin{table}[!ht]
	\centering
	\renewcommand\arraystretch{1.5}{
		\begin{tabular}{ccccc}
			\hline
			\textbf{$\mathsf{K}_0$} & \textbf{$\mathbb{T}$} & \textbf{$\mathbb{T}_{\text{Han}}$} & \textbf{$\mathbb{T}_{\text{Lei, boot}}$} & \textbf{$\mathbb{T}_{\text{Hu, boot}}^{\text{aug}}$} \\ \hline
			\textbf{1} & Reject & Reject & Reject & Reject \\
			\textbf{2} & \bf{Accept} & \bf{Accept} & Reject &  \bf{Accept}  \\ 
			\hline
	\end{tabular}}
	\caption{Testing results for political blog  network.}
\label{political}
	
\end{table}

	\subsection{Sina Weibo data}\label{subsec-sina}

	\noindent
	The second dataset, analyzed by \cite{wu2022inward}, consists of friendship links among 2,580 Sina Weibo users and has an edge density of approximately 0.41\%. In \cite{wu2022inward}, the adjacency matrix $A$ is defined for a directed network, where $A_{ij}=1$ if user $i$ follows user $j$. Using a segmentation procedure, \cite{wu2022inward} show that users can be categorized into four quadrants based on two directional nodal influence indices: “inward influence” (receptivity to influence) and “outward influence” (influence exerted on others).

To fit our undirected network framework, we redefine $A_{ij}=1$ if users $i$ and $j$ follow each other. In this network, the two directional influence indices are naturally combined into a single undirected mutual influence index for each pair of nodes. Applying the segmentation procedure of \cite{wu2022inward} yields two communities. Hence, the true number of communities in this network is taken to be $\mathsf{K}=2$.

We report the results for testing $\mathsf{K}_0=1$ and $2$, since all tests with $\mathsf{K}_0>2$ lead to the same conclusion. As shown in Table~\ref{sinaweibo}, $\mathbb{T}$ rejects $\mathbf{H}_0: \mathsf{K}=1$ but does not reject $\mathbf{H}_0: \mathsf{K}=2$, correctly identifying at least two communities. In contrast, $\mathbb{T}_{\text{Lei, boot}}$, $\mathbb{T}_{\text{Hu,boot}}^{\text{aug}}$, and $\mathbb{T}_{\text{Han}}$ reject all null hypotheses.

Therefore, the testing results of $\mathbb{T}$ are consistent with the ground truth $\mathsf{K}=2$, demonstrating the effectiveness of our method in  sparse networks, whereas the other methods fail to correctly identify the community structure.

\begin{table}[!ht]
	\centering
	\renewcommand\arraystretch{1.5}{
		\begin{tabular}{ccccc}
			\hline
			\textbf{$\mathsf{K}_0$} & \textbf{$\mathbb{T}$} & \textbf{$\mathbb{T}_{\text{Han}}$} & \textbf{$\mathbb{T}_{\text{Lei, boot}}$} & \textbf{$\mathbb{T}_{\text{Hu, boot}}^{\text{aug}}$} \\ \hline
			\textbf{1} & Reject & Reject & Reject & Reject \\
			\textbf{2} & \bf{Accept} &Reject & Reject &  Reject \\ 
			\hline
	\end{tabular}}
	\caption{Testing results for Sina Weibo  network.}
\label{sinaweibo}
	
\end{table}


	{		
	\section{Concluding remarks}\label{sec-conclude}
	
	\noindent
In this paper, we propose a model-free spectral method for testing the number of communities in networks. The proposed procedure is computationally efficient and easy to implement. Compared with existing approaches, our test does not require explicit model fitting and demonstrates clear advantages in accommodating sparse networks and a diverging number of communities. In particular, we establish an explicit trade-off between network sparsity and the number of communities, namely $
n^{1/3} \max_{i,j} P_{ij} / \mathsf{K}^2 \rightarrow \infty.$ We show that, under suitable regularity conditions, the proposed test statistic converges to a functional of the type-I Tracy--Widom distribution characterized by the Airy kernel under the null hypothesis. We further derive the asymptotic power of the test and establish the consistency of the proposed sequential estimators.

Three promising directions for future research remain. First, extending our method to extremely sparse networks \citep{han2023universal} would be of considerable theoretical interest. Second, generalizing the proposed framework to accommodate nonparametric graphical models \citep{wolfe2013nonparametric} represents another important direction. Third, incorporating correlated binary network data, such as those generated from latent space models \citep{hoff2002latent}, is also worthy of further investigation. We believe these developments will further enhance the applicability of the proposed method.
}

%
%

	
		\bigskip
		\begin{center}
			{\large\bf SUPPLEMENTARY MATERIAL}
		\end{center}

\noindent
In this supplement, we provide detailed technical proofs, supporting lemmas, additional remarks, and further numerical results. Throughout the rest of the paper, we will consistently use the following conventions. For any matrix $B$, denote by $B_{m \times n}$ the submatrix of $B$ consisting of its last $m$ rows and $n$ columns. In addition, let $\lambda_s(B)$ denote the $s$-th largest eigenvalue of $B$, and define $\widetilde{\lambda}_s(B)$ as the $s$-th largest eigenvalue of $B$ in absolute value. Moreover, for a matrix $B$, define $\|B\| = \lambda_1(B^\top B)^{1/2}$, and for a vector $\bm{b}$, define $|\bm{b}| = (\bm{b}^\top \bm{b})^{1/2}$. Recall that for sequences $\{a_n\}$ and $\{b_n\}$ depending on $n$, $a_n=\mathrm{O}(b_n)$ means $|a_n|\le \mathsf{C}|b_n|$ for some constant $\mathsf{C}>0$, while $a_n=\mathrm{o}(b_n)$  means $|a_n|\le c_n|b_n|$ for some positive sequence $c_n\downarrow0$ as $n\to\infty$. We also use  $a_n\asymp b_n$ to mean both $a_n=\mathrm{O}(b_n)$ and $b_n=\mathrm{O}(a_n)$. For two sequences of random variables $\{a_n\}$ and $\{b_n\}$ depending on $n$, we denote $a_n=\mathrm{O}_{\mathbb{P}}(b_n)$ if $a_n=\mathrm{O}(b_n)$ holds with probability $1-\mathrm{o}(1)$. Similarly, we can define $a_n=\mathrm{o}_{\mathbb{P}}(b_n)$ if $a_n=\mathrm{o}(b_n)$ holds with probability $1-\mathrm{o}(1)$.


%

\setcounter{equation}{0}
\setcounter{section}{0}
\setcounter{table}{0}
\setcounter{figure}{0}

\renewcommand{\theequation}{S.\arabic{equation}}
\renewcommand{\thesection}{S.\arabic{section}}
\renewcommand{\thetable}{S.\arabic{table}}
\renewcommand{\thefigure}{S.\arabic{figure}}
\renewcommand{\thealgocf}{S.\arabic{algocf}}
\setcounter{algocf}{0}
\setcounter{theorem}{0}
\setcounter{lemma}{0}
\setcounter{remark}{0}

\renewcommand{\thetheorem}{S.\arabic{theorem}}
\renewcommand{\thelemma}{S.\arabic{lemma}}
\renewcommand{\theremark}{S.\arabic{remark}}
	



\section{Technical proof} \label{apen-sec2}
\noindent
In this section, we present the technical proofs of the results in Section~\ref{sec-property}. Recall the adjacency matrix $A=(A_{ij})\in\mR ^{n\times n}$ and consider
\begin{equation}\label{eq:A1}
	\widetilde A=A-\mathbb{E}(A),
\end{equation}
where $\mathbb{E}(A)$ is the expected value of $A$. Hence, $\widetilde A$ has zero-mean entries and zero diagonal elements.
It is straightforward to see that entries of $\widetilde A$ have finite variances. Given $\mathbb{E}(A)=P-\text{diag}(P)$, (\ref{eq:A1}) can then be
rewritten as
\beq\label{eq:ARP}
A=P+\widetilde A-\text{diag}(P),
\eeq
where $\text{diag}(P)$ represents the diagonal matrix constructed from $(P_{11},\cdots,P_{nn})^\top$.


\subsection{Proof of Theorem~\ref{theory-null}}\label{apen-sec3}
\noindent  The proof of Theorem~\ref{theory-null} can be split into five steps, as follows. For notational simplicity, we denote
\begin{equation*}
 p_{\max} = \max_{i, j} P_{ij}.
\end{equation*}
Recall (\ref{eq:A1}) and (\ref{eq:ARP}). {
Step I shows that, for $1\le s\le \mathsf{K}_{\max}-\mathsf{K}_0+2$, \[\frac{1}{np_{\max}}\lambda_{\mathsf{K}_0+s}(AA^\top)=\frac{1}{np_{\max}}\lambda_s({\widecheck A}_{2}{\widecheck A}_{2}^\top)+\mathrm{o}_{\mathbb{P}}\left(n^{-2/3}\right),\] 
where ${\widecheck A}_{2}$ is a submatrix of $\widetilde A$ after appropriate transformations. Step II demonstrates that
\begin{equation}\label{eq:AR221} 
\frac{1}{\sqrt{np_{\max}}}\lambda_{\mathsf{K}_0+s}(A)= \frac{\xi_s}{\sqrt{np_{\max}}}\lambda_s(\widetilde A_{G_I})+\mathrm{o}_{\mathbb{P}}\left(n^{-2/3}\right),\end{equation}
where $\xi_s>0$ is some constant, and $\widetilde A_{G_I}$ is a submatrix of $\widetilde A$ satisfying $\lambda_s(\widetilde A_{G_I})\asymp \lambda_s({\widecheck A}_{2})$ holds with probability $1-\mathrm{o}(1)$. 
Step III establishes that \[c_{i^*j^*}\widetilde\Omega_{G_I,i^*j^*}\lambda_{\mathsf{K}_0+s}(A)=\lambda_{s}({\widetilde A}_{G_I}\circ\widetilde\Omega_{G_I})+\mathrm{o}_{\mathbb{P}}\left(n^{-2/3}\right),\] where $c_{i^*j^*}$ and $\widetilde\Omega_{G_I,i^*j^*}$ are some deterministic scalars satisfying $c_{i^*j^*}\widetilde\Omega_{G_I,i^*j^*}\asymp\frac{1}{\sqrt{np_{\max}}}$, and ${\widetilde A}_{G_I}\circ\widetilde\Omega_{G_I}$ represents the variance-normalized version of ${\widetilde A}_{G_I}$.
Step IV shows that 
for $1\le s\le \mathsf{K}_{\max}-\mathsf{K}_0+2$ and any $x_s\in\mR$,
\begin{align*}
	&\lim_{n \rightarrow \infty}\mathbb{P}\bigl(n^{2/3}(\lambda_{1}({\widetilde A}_{G_I}\circ\widetilde\Omega_{G_I})-\mathsf{L})\leq x_1,\cdots,
	n^{2/3}(\lambda_{s}({\widetilde A}_{G_I}\circ\widetilde\Omega_{G_I})-\mathsf{L})\leq x_s\bigr) \\
	=&\lim_{n \rightarrow \infty}\mathbb{P}\bigl(n^{2/3}(\lambda_{1}(W)-2)\leq x_1,\cdots, n^{2/3}(\lambda_{s}(W)-2)\leq x_s\bigr), 
\end{align*}
where $W$ is a GOE matrix and $\mathsf{L}$ is some constant. Step V derives that, for any $x\in\mR$,
\[\lim_{n \rightarrow \infty}F_{\mathbb{T}}(x)=\lim_{n \rightarrow \infty} F_{\mathbb{T}_W}(x),\]  
where $F_{\mathbb{T}}(x)$ and $F_{\mathbb{T}_W}(x)$ are the cumulative distribution functions of $\mathbb{T}$ and $\mathbb{T}_W$.
We now provide the details of the proof following the aforementioned five steps.}


\begin{proof}[\bf Proof of Theorem \ref{theory-null}]	
\textbf{Step I.} To bound the diagonal entries of $P$ by $\mathrm{O}(n^{-1/3})$ for the purpose of deriving
	\eqref{eq:AR221}, while keeping $\mathbb{E}(A)=P-\text{diag}(P)$ unchanged to facilitate the technical analysis, we apply a diagonal permutation to $P$. Specifically, let 
\begin{equation*}
P' = P + \Delta,
\end{equation*}
where $\Delta = -\left(1 - n^{-1/3}\right)\text{diag}(P)$ is a diagonal perturbation matrix. 
By Weyl’s inequality, we have
\beq\label{PP:Weyl}
|\lambda_i(P') - \lambda_i(P)| \le \|\Delta\| = \max_i \left(1 - n^{-1/3}\right)|P_{ii}|.
\eeq
Under the conditions $n^{1/3}p_{\max}/\mathsf{K}^2\to\infty$ and  $\min_{1\le k\le \mathsf{K}}|\lambda_{k}(P)|\ge cnp_{\max}/\mathsf{K}$ for some constant $c>0$, we can derive that $\min_{1 \le k \le \mathsf{K}} |\lambda_k(P)| \ge cn^{2/3}$. Thus,
\[\min_{1 \le k \le \mathsf{K}} |\lambda_k(P)| \gg \|\Delta\|.\] 
This implies that $\min_{1 \le k \le \mathsf{K}} |\lambda_k(P')| \gg \|\Delta\|$ and no eigenvalue of $P'$ crosses zero.
Therefore, this diagonal perturbation introduces a negligible change in the leading eigenvalues of $P$, making the rank of $P'$ equal to that of $P$ for sufficiently large $n$.

Based on this perturbation, it is straightforward to verify that $\mathbb{E}(A) = P' - \text{diag}(P')$. Recall \eqref{eq:A1} that $\widetilde A=A-\mathbb{E}(A)$. We can rewrite \eqref{eq:ARP} as
\beq A = P' + \widetilde{A} - \operatorname{diag}(P'),\label{eq:ARP'}
\eeq
where $A$, $P'$, and $\widetilde A-\text{diag}(P')$ are symmetric matrices. 
Under $\mathbf{H}_0$, the rank of $P'$ is $\mathsf{K}_0$.  Let $P'_*=HP'$, where $H$ is an orthogonal matrix used to perform an orthogonal transformation on $P'$ such that the first $\mathsf{K}_0$ columns of $P'_*$ are nonzero and the remaining $n-\mathsf{K}_0$ columns of $P'_*$ are zero. Denote 
$A_*=HA$ and $\widetilde A_*=H(\widetilde A-\text{diag}(P'))$. It then follows from \eqref{eq:ARP'} that
\beq \label{eq:Astar}
A_*=P'_*+\widetilde A_*.
\eeq 
{By Lemma~\ref{lemma:invariance_eigen}, the eigenvalues of $A_*A_*^{\top}$, $P'_*P_*^{'\top}$, and $\widetilde A_*\widetilde A_*^{\top}$ are the same as those of $AA^\top$, $P'P^{'\top}$, and $(\widetilde A-\text{diag}(P'))(\widetilde A-\text{diag}(P'))^\top$, respectively. }  

Denote the matrix consisting of the
{$\mathsf{K}_0$ columns of $P'_*$ associated with the nonzero eigenvalues  as $P'_{*,1}\in\mR^{n\times \mathsf{K}_0}$, the matrix of the first $\mathsf{K}_0$ columns of $\widetilde A_*$ as $\widetilde A_{*,1}\in\mR^{n\times \mathsf{K}_0}$ and the matrix of the last $n-\mathsf{K}_0$ columns of $\widetilde A_*$ as $\widetilde A_{*,2}\in\mR^{n\times (n-\mathsf{K}_0)}$.} Then, based on \eqref{eq:Astar}, we can decompose $A_*A_*^\top$ into a sum of two terms given below,
\beq \label{eq:decom}
A_*A_*^\top=(P'_{*,1}+\widetilde A_{*,1})(P'_{*,1}+\widetilde A_{*,1})^\top+\widetilde A_{*,2}\widetilde A_{*,2}^\top.
\eeq
\noindent Let $G^\top\Lambda_{P'_{*,1}\widetilde A_{*,1}} G$ be a spectral decomposition of the first term $(P'_{*,1}+\widetilde A_{*,1})(P'_{*,1}+\widetilde A_{*,1})^\top$, where $G$ is an orthogonal matrix. Since the rank of $(P'_{*,1}+\widetilde A_{*,1})(P'_{*,1}+\widetilde A_{*,1})^\top$ is  $\mathsf{K}_0$,  the diagonal matrix $\Lambda_{P'_{*,1}\widetilde A_{*,1}}$ can be chosen such that the first {$\mathsf{K}_0$ diagonal elements are nonzero and the remaining $n-\mathsf{K}_0$} diagonal elements are zero, i.e., 
\[\Lambda_{P'_{*,1}\widetilde A_{*,1}}=\text{diag}\left\{\lambda_1\big((P'_{*,1}+\widetilde A_{*,1})(P'_{*,1}+\widetilde A_{*,1})^\top\big),\cdots,\lambda_{\mathsf{K}_0}\big((P'_{*,1}+\widetilde A_{*,1})(P'_{*,1}+\widetilde A_{*,1})^\top\big),0,\cdots,0\right\}.\]
In addition, let $\widecheck A=G\widetilde A_{*,2}$,  and denote the matrices consisting of the first $\mathsf{K}_0$ rows and the last $n-\mathsf{K}_0$ rows of  $\widecheck A$ as $\widecheck A_1\in\mR^{\mathsf{K}_0\times (n-\mathsf{K}_0)}$ and $\widecheck A_2\in\mR^{ (n-\mathsf{K}_0)\times (n-\mathsf{K}_0)}$, respectively. Then, by \eqref{eq:decom}, we have
\[\frac{1}{np_{\max}}GA_*A_*^\top G^\top=\frac{1}{np_{\max}}\Lambda_{P'_{*,1}\widetilde A_{*,1}}+\frac{1}{np_{\max}}\widecheck A\widecheck A^\top.\]
Define $\widetilde d=\frac{\lambda_{\mathsf{K}_0}\big((P'_{*,1}+\widetilde A_{*,1})(P'_{*,1}+\widetilde A_{*,1})^\top\big)}{2np_{\max}}$.  According to Lemma~\ref{lemma:bound}, for finite $1\le s\le \mathsf{K}_{\max}-\mathsf{K}_0+2$,
we have
\beq\label{eq:null-inequa}
\Big|\lambda_{\mathsf{K}_0+s}\Bigl(\frac{1}{np_{\max}}GA_*A_*^\top G^\top\Bigr)-\frac{1}{np_{\max}}\lambda_s(\widecheck A_2\widecheck A_2^\top)\Big|\le 3\widetilde d\frac{n^{-2}p_{\max}^{-2}\|{\widecheck A}_{2}{\widecheck A}_{2}^\top\|^2}{(\widetilde d-n^{-1}p_{\max}^{-1}\|{\widecheck A}_{2}{\widecheck A}_{2}^\top\|)^2}. \;\; 
\eeq

We next control the right-hand side of (\ref{eq:null-inequa}).  
Let
\[G_{I}=\begin{pmatrix} 0_{\mathsf{K}_0\times \mathsf{K}_0} & 0_{\mathsf{K}_0\times (n-\mathsf{K}_0)}\\ 0_{(n-\mathsf{K}_0)\times \mathsf{K}_0} & I_{(n-\mathsf{K}_0)} \end{pmatrix},\]
where $I_{m}$ represents the $m$-dimensional identity matrix. Then,  ${\widecheck A}_{2}$  can be re-expressed as
\beq\label{eq:A2check}
{\widecheck A}_{2}=\{G_{I}GH(\widetilde A-\text{diag}(P'))G_I\}_{(n-\mathsf{K}_0)\times (n-\mathsf{K}_0)}.
\eeq
{Under the condition $n^{1/3}p_{\max}/\mathsf{K}^2\to\infty$, we have $\lim_{n\to\infty}np_{\max}\log^{-1}(n)\to\infty$. Then, by Lemma~\ref{lemma:sigular_bound}, there exists some constant $\psi_{\widetilde A}>0$ such that with probability $1-\mathrm{o}(1)$ }
\beq \label{eq:R}
\|\widetilde A\|\equiv\widetilde\lambda_1(\widetilde A) \leq \psi_{\widetilde A}\sqrt{np_{\max}}.
\eeq

\noindent This result, together with \eqref{eq:A2check} and the fact that both $H$ and $G$ are orthogonal, implies that 
\begin{align}\label{eq:A22}
\|{\widecheck A}_{2}{\widecheck A}_{2}^\top\|&\le\|G_I\|\|(\widetilde A-\text{diag}(P'))(\widetilde A-\text{diag}(P'))^\top\| \\
&\le\|{\widetilde A}{\widetilde A}^\top\|+2\|{\widetilde A}\|\|\text{diag}(P')\|+\|\text{diag}^2(P')\| \nonumber \\
&\le\widetilde\lambda_1^2({\widetilde A})+2\widetilde\lambda_1({\widetilde A})p_{\max}+p_{\max}^2 \nonumber\\
& =\mathrm{O}_\mathbb{P}(np_{\max}). \nonumber
\end{align}
This implies that, there exists some constant $\psi_{\widecheck A_{2}}>0$ such that with probability $1-\mathrm{o}(1),$ $\|{\widecheck A}_{2}{\widecheck A}_{2}^\top\| \leq  \psi_{\widecheck A_{2}}np_{\max}$.

In addition, let $$c=\widetilde d/\psi_{\widecheck A_{2}}^2-2/\psi_{\widecheck A_{2}}+1/\widetilde d.$$ After straightforward calculations, we  obtain that the right-hand side of (\ref{eq:null-inequa}) is smaller than $3c^{-1}$ with probability $1-\mathrm{o}(1)$. {We next demonstrate that $c^{-1}= \mathrm{o}_\mathbb{P}(n^{-2/3})$.}
Applying techniques analogous to those used in deriving \eqref{eq:A22},  for some constant $\psi_{{\widetilde A}_1}>0$, we can obtain that with probability $1-\mathrm{o}(1)$
\beq\label{eq:R*1}
\| {\widetilde A}_{*,1} {\widetilde A}_{*,1}^\top\|\le\|({\widetilde A}-\text{diag}(P'))({\widetilde A}-\text{diag}(P'))^\top\| \leq \psi_{{\widetilde A}_1}np_{\max}.
\eeq
Note that $P'_{ij}\leq p_{\max}$ by definition, we then have $\widetilde\lambda_1^2(P’)\leq \mbox{tr}(P^{'\top} P')=\sum_{i,j}P_{ij}^{'2}\leq n^2 p_{\max}^2$, which leads to
$\widetilde\lambda_1(P')\leq np_{\max}$. Denote $\bm\alpha_{\mathsf{K}_0}$ be the unit eigenvector corresponding to the $\mathsf{K}_0$-th largest eigenvalue of $(P'_{*,1}+{\widetilde A}_{*,1})(P'_{*,1}+{\widetilde A}_{*,1})^\top$.
The above result,
together with \eqref{PP:Weyl},  \eqref{eq:R}, \eqref{eq:R*1} and the assumption $\min_{1\le k\le \mathsf{K}}|\lambda_{k}(P)|\ge cnp_{\max}/\mathsf{K}$,  implies that, with probability $1-\mathrm{o}(1),$ we have that 
\begin{align*}
&\|\lambda_{\mathsf{K}_0}\big((P'_{*,1}+{\widetilde A}_{*,1})(P'_{*,1}+{\widetilde A}_{*,1})^\top\big)\bm\alpha_{\mathsf{K}_0}\|\\
&\ge\|P'_{*,1}P_{*,1}^{'\top}\bm\alpha_{K_0}\|-\|P'_{*,1}{\widetilde A}_{*,1}^\top\|\|\bm\alpha_{K_0}\|-\|{\widetilde A}_{*,1}P_{*,1}^{'\top}\|\|\bm\alpha_{\mathsf{K}_0}\|-\|{\widetilde A}_{*,1}{\widetilde A}_{*,1}^\top\|\|\bm\alpha_{\mathsf{K}_0}\|\\
&\geq \|P'_{*,1}P_{*,1}^{'\top}\bm\alpha_{\mathsf{K}_0}\|-2\|P'\|\|{\widetilde A}\|-2\|P'\|\|\text{diag}(P')\|-\psi_{{\widetilde A}_1}np_{\max}\\
&\geq \|P'_{*,1}P_{*,1}^{'\top}\bm\alpha_{\mathsf{K}_0}\|-2\widetilde\lambda_1(P')\widetilde\lambda_1({\widetilde A})-2\widetilde\lambda_1(P')\max_iP'_{ii}-\psi_{{\widetilde A}_1}np_{\max}\\
&\geq c^*_{\mathsf{K}_0}\widetilde\lambda_{\mathsf{K}_0}^2(P')-2\psi_{{\widetilde A}}np_{\max}\sqrt{np_{\max}}-2c_4^*np^2_{\max}-\psi_{{\widetilde A}_1}np_{\max} \\
&\geq c^*_{\mathsf{K}_0}n^2p_{\max}^2\mathsf{K}_0^{-2}-2\psi_{{\widetilde A}}n^{3/2}p_{\max}^{3/2}-2c_4^*np_{\max}^2-\psi_{{\widetilde A}_1}np_{\max}\\
&\geq \widetilde c^*_{\mathsf{K}_0}n^2p_{\max}^2\mathsf{K}_0^{-2},\stepcounter{equation}\tag{\theequation}\label{eq:1amPR}
\end{align*}
where $\widetilde c^*_{\mathsf{K}_0}$, $c^*_{\mathsf{K}_0}$, $c_4^*$, $\psi_{{\widetilde A}}$ and $\psi_{{\widetilde A}_1}$ are some positive constants. Consequently, with probability $1-\mathrm{o}(1)$,
\[\widetilde d=\frac{\lambda_{\mathsf{K}_0}\big((P'_{*,1}+{\widetilde A}_{*,1})(P'_{*,1}+{\widetilde A}_{*,1})^\top\big)}{2np_{\max}} \geq \frac{\widetilde c^*_{\mathsf{K}_0}np_{\max}\mathsf{K}_0^{-2}}{2}.\] 
Recall $c=\widetilde d/\psi_{\widecheck A_{2}}^2-2/\psi_{\widecheck A_{2}}+1/\widetilde d$. Then it follows that $c^{-1}=\mathrm{O}_\mathbb{P}\left(n^{-1}p^{-1}_{\max}\mathsf{K}_0^2\right)$. Together with the condition $n^{1/3}p_{\max}/\mathsf{K}_0^2\to\infty$, we have $c^{-1}= \mathrm{o}_\mathbb{P}(n^{-2/3})$.
Recall that the right-hand side of \eqref{eq:null-inequa} is smaller than $3c^{-1}$. It then follows from \eqref{eq:null-inequa} that
 \[\frac{1}{np_{\max}}\lambda_{\mathsf{K}_0+s}(GA_*A_*^\top G^\top)=\frac{1}{np_{\max}}\lambda_s({\widecheck A}_{2}{\widecheck A}_{2}^\top)+\mathrm{o}_\mathbb{P}(n^{-2/3}).\]
{Moreover, since $G$ and $H$ are orthogonal matrices, by Lemma~\ref{lemma:invariance_eigen}, we have}
\beq\label{eq:AA-R22*R22*}
\frac{1}{np_{\max}}\lambda_{\mathsf{K}_0+s}(AA^\top)=\frac{1}{np_{\max}}\lambda_s({\widecheck A}_{2}{\widecheck A}_{2}^\top)+\mathrm{o}_\mathbb{P}\left(n^{-2/3}\right),
\eeq
which completes the first step of the proof.

\noindent \textbf{Step II.}
Recall from \eqref{eq:A2check} that ${\widecheck A}_{2}=\{G_{I}GH({\widetilde A}-\text{diag}(P'))G_I\}_{(n-\mathsf{K}_0)\times (n-\mathsf{K}_0)}$.
To verify the result of this step, we introduce an auxiliary matrix
\beq\label{eq:tildeA2}
{\widetilde A}_{2}=\{G_{I}({\widetilde A}-\text{diag}(P'))G_I\}_{(n-\mathsf{K}_0)\times (n-\mathsf{K}_0)}.
\eeq
Since $H$ and $G$ are orthogonal,  the matrices $GH({\widetilde A}-\text{diag}(P'))G_I({\widetilde A}-\text{diag}(P'))H^\top G^\top$ and $({\widetilde A}-\text{diag}(P'))G_I({\widetilde A}-\text{diag}(P'))$ have the same  eigenvalues $\lambda_{{\widetilde A},1},\cdots,\lambda_{{\widetilde A},n}$. As a result, their corresponding spectral decompositions are
\begin{center}
$GH({\widetilde A}-\text{diag}(P'))G_I({\widetilde A}-\text{diag}(P'))H^\top G^\top=S_1^\top\Lambda_{{\widetilde A}}S_1,$ \\
$({\widetilde A}-\text{diag}(P'))G_I({\widetilde A}-\text{diag}(P'))=S_2^\top\Lambda_{{\widetilde A}}S_2,$
\end{center}
where $\Lambda_{{\widetilde A}}=\text{diag}\{\lambda_{{\widetilde A},1},\cdots,\lambda_{{\widetilde A},n}\}$, and $S_1$ and $S_2$ are the orthogonal matrices associated to these two decompositions. These lead to
\beq\label{eq:R*22}
{\widecheck A}_{2}{\widecheck A}_{2}^\top=\{G_IS_1^\top\Lambda_{{\widetilde A}}S_1G_I\}_{(n-\mathsf{K}_0)\times(n-\mathsf{K}_0)} ~~  \mbox{and} ~~{\widetilde A}_{2} {\widetilde A}_{2}^\top=\{G_IS_2^\top\Lambda_{{\widetilde A}}S_2G_I\}_{(n-\mathsf{K}_0)\times(n-\mathsf{K}_0)}.
\eeq
Using the fact that $S_1$ and $S_2$ are orthogonal, equation (\ref{eq:R*22}) implies that for $1\le s\le \mathsf{K}_{\max}-\mathsf{K}_0+2$, there exist some constants $\xi_s>0$ such that
\[\frac{1}{np_{\max}}\lambda_s({\widecheck A}_{2}{\widecheck A}_{2}^\top)=\frac{\xi_s^2}{np_{\max}}\lambda_s({\widetilde A}_{2} {\widetilde A}_{2}^\top). \] 
This, together with Step I's result \eqref{eq:AA-R22*R22*}, implies
\[\frac{1}{np_{\max}}\lambda_{\mathsf{K}_0+s}(AA^\top)=\frac{\xi_s^2}{np_{\max}}\lambda_s({\widetilde A}_{2}{\widetilde A}_{2}^\top)+\mathrm{o}_\mathbb{P}\left(n^{-2/3}\right).\]

Recall for any square matrix $B,$ we denote $\widetilde{\lambda}_s(B)$ as the $s$-th largest eigenvalue of $B$ in absolute value. Since $A$ and $\widetilde{A}_2$ are symmetric, their singular values coincide with their absolute eigenvalues. Hence, we have
\[\frac{1}{np_{\max}}\widetilde\lambda^2_{\mathsf{K}_0+s}(A)=\frac{\xi_s^2}{np_{\max}}\widetilde\lambda^2_s({\widetilde A}_{2})+\mathrm{o}_\mathbb{P}\left(n^{-2/3}\right).\]

{
Under the conditions that $n^{1/3}p_{\max}/\mathsf{K}^2\to\infty$, $\min_{1\le k\le \mathsf{K}}|\lambda_{k}(P)|\ge cnp_{\max}/\mathsf{K}$ for some constant $c>0$, the assumption that $\max_{ij}P_{ij} \asymp \min_{ij} P_{ij}$, and $\mathsf{K}_{\max}-\mathsf{K}_0$ is finite, Lemma~\ref{lemma:eigsA_cards} implies that $A$ has at least $\mathsf{K}_{\max}+2$ eigenvalues bounded away from zero, which yields that
\[\frac{1}{\sqrt{np_{\max}}}\widetilde\lambda_{\mathsf{K}_0+s}(A)=\frac{\xi_s}{\sqrt{np_{\max}}}\widetilde\lambda_s({\widetilde A}_{2})+\mathrm{o}_\mathbb{P}\left(n^{-2/3}\right).\] 
This implies that, there exists some $\psi\in\{1,-1\}$ such that 
\[\frac{1}{\sqrt{np_{\max}}}\lambda_{\mathsf{K}_0+s}(A)=\psi\frac{\xi_s}{\sqrt{np_{\max}}}\lambda_s({\widetilde A}_{2})+\mathrm{o}_\mathbb{P}\left(n^{-2/3}\right).\]
Since the choice of  $\psi$ leads to the same subsequent asymptotic arguments, we take $\psi=1$ without loss of generality and obtain }
\beq\label{eq:A-R22}
\frac{1}{\sqrt{np_{\max}}}\lambda_{\mathsf{K}_0+s}(A)=\frac{\xi_s}{\sqrt{np_{\max}}}\lambda_s({\widetilde A}_{2})+\mathrm{o}_\mathbb{P}\left(n^{-2/3}\right).
\eeq

%

Recall from \eqref{eq:tildeA2} that ${\widetilde A}_{2}=\{G_I({\widetilde A}-\text{diag}(P'))G_I\}_{(n-\mathsf{K}_0)\times (n-\mathsf{K}_0)}$. By the definition of $P'$, we have
\beq\label{eq:GI}
G_I{\widetilde A}G_I-\max_iP'_{ii}G_I\le G_I{\widetilde A}G_I-G_I\text{diag}(P')G_I\le G_I{\widetilde A}G_I-\min_iP'_{ii}G_I.
\eeq
Subsequently, we define 
\beq\label{eq:AGI}
{\widetilde A}_{G_I}=\{G_{I}{\widetilde A}G_I\}_{(n-\mathsf{K}_0)\times (n-\mathsf{K}_0)}.
\eeq
Employing \eqref{eq:GI} and noting that all entries $P'_{ii}$ are of the same order, we can obtain
\beq\label{eq:A2-AGI}
\lambda_s({\widetilde A}_{2})=\lambda_s({\widetilde A}_{G_I})-\xi_s^*\max_iP'_{ii},
\eeq
where, $\xi_s^*>0$  are some constants for $1\le s\le \mathsf{K}_{\max}-\mathsf{K}_0+2$. Since $P'_{ii}=\mathrm{O}(n^{-1/3})$ by definition, under the condition $n^{1/3}p_{\max}/\mathsf{K}^2\to\infty$, we have $\frac{\max_i P'_{ii}}{\sqrt{n p_{\max}}} = \mathrm{o}\left(n^{-2/3}\right)$.
This, in conjunction with \eqref{eq:A-R22} and \eqref{eq:A2-AGI}, yields
\beq\label{eq:AR22}
\frac{1}{\sqrt{np_{\max}}}\lambda_{\mathsf{K}_0+s}(A)=\frac{\xi_s}{\sqrt{np_{\max}}}\lambda_s({\widetilde A}_{G_I})+\mathrm{o}_\mathbb{P}\left(n^{-2/3}\right),
\eeq
\noindent which completes the second step of the proof.

\noindent \textbf{Step III.} 
Define a symmetric matrix $\Omega=(\Omega_{ij})\in\mR^{n\times n}$ where $\Omega_{ij}=\{P_{ij}(1-P_{ij})\}^{-1/2}$ for $1\le i, j\le n$. Furthermore, define  
\beq\label{eq:OmegaGI}
\widetilde\Omega_{G_I}=(n-\mathsf{K}_0-1)^{-1/2}\{G_I\Omega G_I\}_{(n-\mathsf{K}_0)\times(n-\mathsf{K}_0)}.
\eeq
As all entries $P_{ij}$ are of the same order, it implies that $\widetilde\Omega_{G_I,ij}\asymp\frac{1}{\sqrt{n p_{\max}}}$ for $1\le i,j\le n-\mathsf{K}_0$. This, together with \eqref{eq:AR22}, implies that 
\beq\label{eq:AR22-n-k0}
\widetilde\Omega_{G_I,ij}\lambda_{\mathsf{K}_0+s}(A)= \xi_s\widetilde\Omega_{G_I,ij}\lambda_s({\widetilde A}_{G_I})+\mathrm{o}_\mathbb{P}\left(n^{-2/3}\right).
\eeq 
{Let $\circ$ denote the Hadamard (entrywise) product. Applying Lemma~\ref{lemma:hadamard_eigen}, we have}
\[\lambda_s(\widetilde\Omega_{G_I}\circ {\widetilde A}_{G_I}\widetilde A^\top_{G_I}\circ\widetilde\Omega_{G_I})\geq \min_{i}\widetilde\Omega^2_{G_I,ii} \lambda_s({\widetilde A}_{G_I}\widetilde A^\top_{G_I}),\]
\[\lambda_s(\widetilde\Omega_{G_I}\circ{\widetilde A}_{G_I}\widetilde A^\top_{G_I}\circ\widetilde\Omega_{G_I})\leq \max_{i}\widetilde\Omega^2_{G_I,ii}\lambda_s({\widetilde A}_{G_I}{\widetilde A}^\top_{G_I}). \]
Since all entries of $\widetilde\Omega_{G_I}$ are of the same order by definition, for any fixed pair $(i^*, j^*)$, there exist some constants $0<m_1\leq m_2<\infty$  independent of $s$ and $n$ such that
\[m_1^2\xi_s^2\widetilde\Omega^2_{G_I,i^* j^*} \lambda_s({\widetilde A}_{G_I}\widetilde A^\top_{G_I})\leq \lambda_s\Big(\widetilde\Omega_{G_I}\circ{\widetilde A}_{G_I}\widetilde A^\top_{G_I}\circ\widetilde\Omega_{G_I}\Big)\leq m_2^2\xi_s^2\widetilde\Omega^2_{G_I,i^* j^*} \lambda_s({\widetilde A}_{G_I}\widetilde A^\top_{G_I}),\]
where $\xi_s>0$ is some constant as in \eqref{eq:A-R22}, and the constants $m_1$ and $m_2$ ensure that the entries of $\xi_s\widetilde{\Omega}_{G_I}$ are uniformly bounded for $1 \leq i, j \leq n-\mathsf{K}_0$ and finite $1\le s\le \mathsf{K}_{\max}-\mathsf{K}_0+2$.
By an argument analogous to that used to derive \eqref{eq:A-R22},  we have
\[m_1\xi_s\widetilde\Omega_{G_I,i^* j^*} \lambda_s({\widetilde A}_{G_I})\leq \lambda_s\Big({\widetilde A}_{G_I}\circ\widetilde\Omega_{G_I}\Big)\leq m_2\xi_s\widetilde\Omega_{G_I,i^* j^*} \lambda_s({\widetilde A}_{G_I}).\]
Therefore, for the given $(i^*, j^*)$, there exists some constant $m_1\leq c_{i^*j^*}\leq m_2$ independent of $s$ and $n$ such that
\[ c_{i^*j^*}\xi_s\widetilde\Omega_{G_I,i^*j^*}\lambda_s({\widetilde A}_{G_I})=\lambda_s\Big({\widetilde A}_{G_I}\circ\widetilde\Omega_{G_I}\Big).\]
This result, combined with \eqref{eq:AR22-n-k0}, leads to
\beq\label{eq:AR22-star}
c_{i^*j^*}\widetilde\Omega_{G_I,i^*j^*}\lambda_{\mathsf{K}_0+s}(A)=\lambda_s\Big({\widetilde A}_{G_I}\circ\widetilde\Omega_{G_I}\Big)+\mathrm{o}_\mathbb{P}\left(n^{-2/3}\right).
\eeq
\noindent  Based on \eqref{eq:A1}, \eqref{eq:AGI} and \eqref{eq:OmegaGI}, it is straightforward to verify that ${\widetilde A}_{G_I} \circ \widetilde\Omega_{G_I}$ is the variance-normalized version of ${\widetilde A}_{G_I}$. This completes the third step of the proof.

\noindent \textbf{Step IV.} 
Denote a GOE matrix by $W\in\mR^{n\times n}$, where $W_{ij} = W_{ji}$, the off-diagonal entries $W_{ij}$ are independently distributed as $\mathcal{N}(0, n^{-1})$, and the diagonal entries $W_{ii}$ are distributed as $\mathcal{N}(0, 2/n)$.  It is known that $n^{2/3}(\lambda_{1}(W)-2)$ converges in distribution to the type-I Tracy–Widom distribution \citep{tracy1994level,tracy1996orthogonal}. Moreover, for any fixed $s > 0$, the joint distribution of $\{ n^{2/3}(\lambda_k(W) - 2) \}_{1 \leq k \leq s}$, admits a joint density described by the Airy kernel, also commonly referred to as the Tracy–Widom law \citep{forrester1993spectrum}.

Consider the symmetric matrix ${\widetilde A}_{G_I}\circ\widetilde\Omega_{G_I}$. Under the random graph model \eqref{Ber},  for  $1\le i\not=j\le n-\mathsf{K}_0$,  we can obtain
\begin{center}$\mathbb{E}( {\widetilde A}_{G_I,ij}\circ\widetilde\Omega_{G_I,ij})=0$ and  $\sum_{i}\mathbb{E}| {\widetilde A}_{G_I,ij}\circ\widetilde\Omega_{G_I,ij}|^2=1$.
\end{center}
After some straightforward calculations, we can also verify that, for any $m\ge 2$,
\begin{center}
$\mathbb{E}|{\widetilde A}_{G_I,ij}\circ\widetilde\Omega_{G_I,ij}|^m\le\frac{(cm)^{cm}}{nq^{m-2}}$,
\end{center}
where $c>0$ is some constant and $q=n^{-1/2}(\sum_{i,j}P_{ij})^{1/2}$.
{Under the these moment conditions, together with the assumptions that $n^{1/3}p_{\max}/\mathsf{K}^2\to\infty$ and that $\max_{i,j}P_{ij} \asymp \max_{i,j} P_{ij}$, Lemma~\ref{lemma:hwang} implies that, for any $x_1\in\mR$,
\[\lim_{n \rightarrow \infty}\mathbb{P}\bigl(n^{2/3}(\lambda_{1}({\widetilde A}_{G_I}\circ\widetilde\Omega_{G_I})-\mathsf{L})\leq x_1\bigr)=\lim_{n \rightarrow \infty}\mathbb{P}\bigl(n^{2/3}(\lambda_{1}(W)-2)\leq x_1\bigr),\] 
where $\mathsf{L}$ is a deterministic scaler satisfying $\mathsf{L}-2=\mathrm{o}(1)$ as specified in Lemma~\ref{lemma:hwang}. 
Furthermore, by Lemma~\ref{lemma:joint_hwang}, this convergence can be strengthened to joint convergence.} Specifically, for finite $1\le s\le \mathsf{K}_{\max}-\mathsf{K}_0+2$ and any $x_s\in\mR$,
\begin{align}\label{eq:jointlimit}
	&\lim_{n \rightarrow \infty}\mathbb{P}\bigl(n^{2/3}(\lambda_{1}({\widetilde A}_{G_I}\circ\widetilde\Omega_{G_I})-\mathsf{L})\leq x_1,\cdots,
	n^{2/3}(\lambda_{s}({\widetilde A}_{G_I}\circ\widetilde\Omega_{G_I})-\mathsf{L})\leq x_s\bigr) \\
	=&\lim_{n \rightarrow \infty}\mathbb{P}\bigl(n^{2/3}(\lambda_{1}(W)-2)\leq x_1,\cdots, n^{2/3}(\lambda_{s}(W)-2)\leq x_s\bigr). \nonumber
\end{align}
Hence, the joint distribution of $n^{2/3}(\lambda_{s}({\widetilde A}_{G_I}\circ\widetilde\Omega_{G_I})-\mathsf{L})$ admits a joint density described by the Tracy–Widom law via Airy kernel. This completes the fourth step of the proof.


\noindent \textbf{Step V.}  According to \eqref{eq:AR22-star}, we have
\begin{align*}
\mathbb{T}&=\frac{\lambda_{\mathsf{K}_0+1}(A)-\lambda_{\mathsf{K}_{\max}+1}(A)}{\lambda_{\mathsf{K}_{\max}+1}(A)-\lambda_{\mathsf{K}_{\max}+2}(A)}\\
&=\frac{c_{i^*j^*}\widetilde\Omega_{G_I,i^*j^*}\lambda_{\mathsf{K}_0+1}(A)-c_{i^*j^*}\widetilde\Omega_{G_I,i^*j^*}\lambda_{\mathsf{K}_{\max}+1}(A)}{c_{i^*j^*}\widetilde\Omega_{G_I,i^*j^*}\lambda_{\mathsf{K}_{\max}+1}(A)-c_{i^*j^*}\widetilde\Omega_{G_I,i^*j^*}\lambda_{\mathsf{K}_{\max}+2}(A)}\\
&=\frac{\lambda_{1}({\widetilde A}_{G_I}\circ\widetilde\Omega_{G_I})-\lambda_{\mathsf{K}_{\max}-\mathsf{K}_0+1}({\widetilde A}_{G_I}\circ\widetilde\Omega_{G_I})+\mathrm{o}_{\mathbb{P}}\left(n^{-2/3}\right)}{\lambda_{\mathsf{K}_{\max}-\mathsf{K}_0+1}({\widetilde A}_{G_I}\circ\widetilde\Omega_{G_I})-\lambda_{\mathsf{K}_{\max}-\mathsf{K}_0+2}({\widetilde A}_{G_I}\circ\widetilde\Omega_{G_I})+\mathrm{o}_{\mathbb{P}}\left(n^{-2/3}\right)}\\
&=\frac{n^{2/3}\{\lambda_{1}({\widetilde A}_{G_I}\circ\widetilde\Omega_{G_I})-\mathsf{L}\}-n^{2/3}\{\lambda_{\mathsf{K}_{\max}-\mathsf{K}_0+1}({\widetilde A}_{G_I}\circ\widetilde\Omega_{G_I})-\mathsf{L}\}+\mathrm{o}_\mathbb{P}(1)}{n^{2/3}\{\lambda_{\mathsf{K}_{\max}-\mathsf{K}_0+1}({\widetilde A}_{G_I}\circ\widetilde\Omega_{G_I})-\mathsf{L}\}-n^{2/3}\{\lambda_{\mathsf{K}_{\max}-\mathsf{K}_0+2}({\widetilde A}_{G_I}\circ\widetilde\Omega_{G_I})-\mathsf{L}\}+\mathrm{o}_\mathbb{P}(1)}.
\end{align*}
Recall that
\begin{align*}
\mathbb{T}_W&=\frac{\lambda_{1}(W)-\lambda_{\mathsf{K}_{\max}-\mathsf{K}_0+1}(W)}{\lambda_{\mathsf{K}_{\max}-\mathsf{K}_0+1}(W)-\lambda_{\mathsf{K}_{\max}-\mathsf{K}_0+2}(W)}\\
&=\frac{n^{2/3}\{\lambda_{1}(W)-2\}-n^{2/3}\{\lambda_{\mathsf{K}_{\max}-\mathsf{K}_0+1}(W)-2\}
}{n^{2/3}\{\lambda_{\mathsf{K}_{\max}-\mathsf{K}_0+1}(W)-2\}-n^{2/3}\{\lambda_{\mathsf{K}_{\max}-\mathsf{K}_0+2}(W)-2\}}.
\end{align*}
Consequently, employing \eqref{eq:jointlimit}, we obtain $\lim_{n \rightarrow \infty}F_{\mathbb{T}}(x)=\lim_{n \rightarrow \infty} F_{\mathbb{T}_W}(x)$ for any $x\in\mR$, where  $F_{\mathbb{T}}(x)$ and $F_{\mathbb{T}_W}(x)$ are the cumulative distribution functions of $\mathbb{T}$ and $\mathbb{T}_W$, respectively.
This completes the proof of Theorem~\ref{theory-null}. 
\end{proof}

\subsection{Proof of Theorem~\ref{theory-alter}}\label{apen-sec4}
\noindent	
Under the alternative hypothesis $\mathbf{H}_1: \mathsf{K}_0<\mathsf{K} \leq \mathsf{K}_{\max}$,  we have $\lambda_{\mathsf{K}_0+1}(A)\ge \lambda_{\mathsf{K}}(A)$. This, together with the argument in \eqref{eq:AR22-star}, leads to
\begin{align} \label{eq:alt}
	\mathbb{T}&=\frac{\lambda_{\mathsf{K}_0+1}(A)-\lambda_{\mathsf{K}_{\max}+1}(A)}{\lambda_{\mathsf{K}_{\max}+1}(A)-\lambda_{\mathsf{K}_{\max}+2}(A)} \nonumber \\
	&\ge\frac{\lambda_{\mathsf{K}}(A)-\lambda_{\mathsf{K}_{\max}+1}(A)}{\lambda_{\mathsf{K}_{\max}+1}(A)-\lambda_{\mathsf{K}_{\max}+2}(A)} \nonumber \\ 
	&=\frac{c_{i^*j^*}\widetilde\Omega_{G_I,i^*j^*}
		\lambda_{\mathsf{K}}(A)-c_{i^*j^*}\widetilde\Omega_{G_I,i^*j^*}\lambda_{\mathsf{K}_{\max}+1}(A)}{c_{i^*j^*}\widetilde\Omega_{G_I,i^*j^*}\lambda_{\mathsf{K}_{\max}+1}(A)-c_{i^*j^*}\widetilde\Omega_{G_I,i^*j^*}\lambda_{\mathsf{K}_{\max}+2}(A)}\nonumber \\
	&=\frac{c_{i^*j^*}\widetilde\Omega_{G_I,i^*j^*}\lambda_{\mathsf{K}}(A)-\lambda_{\mathsf{K}_{\max}-\mathsf{K}+1}({\widetilde A}_{G_I}\circ\widetilde\Omega_{G_I})+\mathrm{o}_{\mathbb{P}}\left(n^{-2/3}\right)}{\lambda_{\mathsf{K}_{\max}-\mathsf{K}+1}({\widetilde A}_{G_I}\circ\widetilde\Omega_{G_I})-\lambda_{\mathsf{K}_{\max}-\mathsf{K}+2}({\widetilde A}_{G_I}\circ\widetilde\Omega_{G_I})+\mathrm{o}_{\mathbb{P}}\left(n^{-2/3}\right)}.
\end{align}
{Under the condition $n^{1/3}p_{\max}/\mathsf{K}^2\to\infty$ and applying result (i) of Lemma~\ref{lemma:hwang}, we have
\beq
n^{2/3}\bigl\{\lambda_{\mathsf{K}_{\max}-\mathsf{K}+1}({\widetilde A}_{G_I}\circ\widetilde\Omega_{G_I})-\mathsf{L}\bigr\}=\mathrm{O}_{\mathbb{P}}(n^\phi)~~ \mbox{and} \nonumber
\eeq
\beq \label{eq:eigen}
n^{2/3}\bigl\{\lambda_{\mathsf{K}_{\max}-\mathsf{K}+2}({\widetilde A}_{G_I}\circ\widetilde\Omega_{G_I})-\mathsf{L}\bigr\}=\mathrm{O}_{\mathbb{P}}(n^\phi),
\eeq
for any small $\phi>0$, where $\mathsf{L} \asymp 1$ is a deterministic scalar as specified in Lemma~\ref{lemma:hwang}.  Combining \eqref{eq:alt} and \eqref{eq:eigen}, we have
\begin{align} \label{eq:Tdiverge}
	\mathbb{T}&\ge\frac{\big|c_{i^*j^*}\widetilde\Omega_{G_I,i^*j^*}\lambda_{\mathsf{K}}(A)-\mathsf{L}\big|-\big|\lambda_{\mathsf{K}_{\max}-\mathsf{K}+1}({\widetilde A}_{G_I}\circ\widetilde\Omega_{G_I})-\mathsf{L}\big|+\mathrm{o}_{\mathbb{P}}\left(n^{-2/3}\right)}
	{\bigl\{\lambda_{\mathsf{K}_{\max}-\mathsf{K}+1}({\widetilde A}_{G_I}\circ\widetilde\Omega_{G_I})-\mathsf{L}\bigr\}-\bigl\{\lambda_{\mathsf{K}_{\max}-\mathsf{K}+2}({\widetilde A}_{G_I}\circ\widetilde\Omega_{G_I}-\mathsf{L}\bigr\}+\mathrm{o}_{\mathbb{P}}\left(n^{-2/3}\right)} \nonumber\\
	&=\frac{n^{2/3}\big|c_{i^*j^*}\widetilde\Omega_{G_I,i^*j^*}\lambda_{\mathsf{K}}(A)-\mathsf{L}\big|-n^{2/3}\big|\lambda_{\mathsf{K}_{\max}-\mathsf{K}+1}({\widetilde A}_{G_I}\circ\widetilde\Omega_{G_I})-\mathsf{L}\big|+\mathrm{o}_{\mathbb{P}}(1)}
	{n^{2/3}\bigl\{\lambda_{\mathsf{K}_{\max}-\mathsf{K}+1}({\widetilde A}_{G_I}\circ\widetilde\Omega_{G_I})-\mathsf{L}\bigr\}-n^{2/3}\bigl\{\lambda_{\mathsf{K}_{\max}-\mathsf{K}+2}({\widetilde A}_{G_I}\circ\widetilde\Omega_{G_I}-\mathsf{L}\bigr\}+\mathrm{o}_{\mathbb{P}}(1)}, \nonumber\\
	&=\frac{n^{2/3}\big|c_{i^*j^*}\widetilde\Omega_{G_I,i^*j^*}\lambda_{\mathsf{K}}(A)-\mathsf{L}\big|-\mathrm{O}_{\mathbb{P}}(n^\phi)+\mathrm{o}_{\mathbb{P}}(1)}{\mathrm{O}_{\mathbb{P}}(n^\phi)+\mathrm{o}_{\mathbb{P}}(1)}.
\end{align}
According to \eqref{eq:decom}, \eqref{eq:null-inequa}, and \eqref{eq:A22}, the $\mathsf{K}$ largest eigenvalues of $\frac{1}{np_{\max}} AA^\top$
are asymptotically governed by those of $\frac{1}{np_{\max}}(P'_{*,1}+{\widetilde A}_{*,1})(P'_{*,1}+{\widetilde A}_{*,1})^\top$. According to \eqref{eq:1amPR} and under the condition $n^{1/3}p_{\max}/\mathsf{K}^2\to\infty$,   we have, with probability $1-\mathrm{o}(1)$,
\[\frac{1}{np_{\max}}\lambda_{\mathsf{K}}\big((P'_{*,1}+{\widetilde A}_{*,1})(P'_{*,1}+{\widetilde A}_{*,1})^\top\big)\geq \widetilde c_{\mathsf{K}}^*np_{\max}/\mathsf{K}^2\gg n^{2/3},\]
where $\widetilde c_{\mathsf{K}}^*>0$ is a constant. 
Consequently, we choose $\phi>0$ in \eqref{eq:eigen} such that, with probability $1-\mathrm{o}(1)$, 
\[\frac{1}{np_{\max}}\lambda_{\mathsf{K}}(AA^\top)\gg n^{2\phi}. \] 
Together with the fact that $c_{i^*j^*}\widetilde\Omega_{G_I,i^*j^*}\asymp\frac{1}{\sqrt{np_{\max}}}$,
we have that, with probability $1-\mathrm{o}(1)$, 
\[c_{i^*j^*}\widetilde\Omega_{G_I,i^*j^*}\widetilde\lambda_{\mathsf{K}}(A)\gg n^\phi.\] 
Hence, with probability $1-\mathrm{o}(1)$, 
\beq  \label{eq:abseigen}
\big|c_{i^*j^*}\widetilde\Omega_{G_I,i^*j^*}\lambda_{\mathsf{K}}(A)-\mathsf{L}\big|\gg n^\phi.
\eeq
Combining \eqref{eq:Tdiverge} and \eqref{eq:abseigen}, for any arbitrarily large constant $\mathsf{C}>0$, we have
$\mathbb{P}(\mathbb{T}> \mathsf{C} n^{2/3})=1-\mathrm{o}(1)$ as $n\rightarrow \infty$, which completes the proof.
}

\subsection{Proof of Corollary \ref{cor:hatK-consistency}  }

\noindent 
Consider the estimator $\widehat{\mathsf{K}} = \min \bigl\{ \mathsf{K}_0 \in \{1, \dots, \mathsf{K}_{\max}\} : \mathbb{T} \le c_\alpha \bigr\}$,
where $c_\alpha$ is the upper $\alpha$-quantile of $\mathbb{T}_W$.

For $\mathsf{K}_0 < \mathsf{K}$, Theorem~\ref{theory-alter} implies $\mathbb{P}(\mathbb{T} > c_\alpha) =1-\mathrm{o}(1)$ as  $n \to \infty$.	Hence, all $\mathsf{K}_0 < \mathsf{K}$ are rejected with probability tending to one.

For $\mathsf{K}_0 \geq \mathsf{K}$, Theorem~\ref{theory-null} shows that $\mathbb{T}$ converges in distribution to $\mathbb{T}_W$, so that	$\mathbb{P}(\mathbb{T} \le c_\alpha) = 1 - \alpha-\mathrm{o}(1)$ as  $n \to \infty$. Thus, all $\mathsf{K}_0 \geq \mathsf{K}$ are not rejected with probability tending to $1-\alpha$.

Consequently, since $\widehat{\mathsf{K}}$ is defined as the smallest $\mathsf{K}_0$ not rejected, combining the two cases above gives
\[
\mathbb{P}(\widehat{\mathsf{K}} = \mathsf{K}) = 1 - \alpha-\mathrm{o}(1)  \quad \text{as } n \to \infty,
\]
which completes the proof.

\subsection{Proof of Corollary   \ref{cor:Kstar-consistency} }

\noindent
Consider the estimator  $\widehat{\mathsf{K}}^* = \min \big\{ \mathsf{K}_0 \in \{1,\ldots,\mathsf{K}_{\max}\} : \mathbb{T} \le t_n \big\}$, where $t_n \asymp n^{\epsilon}$ for some $\epsilon \in (0,5/6)$. We prove this corollary by considering two cases, namely underestimation and overestimation.

\begin{itemize}

\item [(i)] \textbf{Underestimation.} For any $\mathsf{K}_0 < \mathsf{K}$, without loss of generality, we assume that $ \lambda_{\mathsf{K}_0+1}(A)\geq \lambda_{\mathsf{K}}(A)\geq  \lambda_{\mathsf{K}_{\max}+1}(A) \geq  \lambda_{\mathsf{K}_{\max}+2}(A) > 0$,
since the alternative case $\lambda_{\mathsf{K}_{\max}+2}(A)\leq \lambda_{\mathsf{K}_{\max}+1}(A)\le \lambda_{\mathsf{K}}(A)\le  \lambda_{\mathsf{K}_0+1}(A) \le 0$ leads to the same conclusion. 

By  \eqref{eq:AR22-star}  and \eqref{eq:eigen}, for the deterministic scalar $c_{i^* j^*}\widetilde\Omega_{G_I,i^* j^*}$ and any  arbitrarily small $\phi>0$, we have 
\[c_{i^*j^*}\widetilde\Omega_{G_I,i^*j^*}\lambda_{\mathsf{K}_{\max}+1}(A)=\mathrm{O}_{\mathbb{P}}(1),\]
\[c_{i^*j^*}\widetilde\Omega_{G_I,i^*j^*}\bigr\{\lambda_{\mathsf{K}_{\max}+1}(A)-\lambda_{\mathsf{K}_{\max}+2}(A)\bigr\}=\mathrm{O}_{\mathbb{P}}(n^{-2/3+\phi}).\]
Since by definition $c_{i^* j^*} \widetilde\Omega_{G_I,i^* j^*} \asymp n^{-1/2}p^{-1/2}_{\max}$, it follows that
\beq\nonumber
\lambda_{\mathsf{K}_{\max}+1}(A)=\mathrm{O}_{\mathbb{P}}(n^{1/2}p^{1/2}_{\max}),
\eeq
\beq\label{eq:eigen_bound}
\lambda_{\mathsf{K}_{\max}+1}(A)-\lambda_{\mathsf{K}_{\max}+2}(A)=\mathrm{O}_{\mathbb{P}}(n^{-1/6+\phi}p^{1/2}_{\max}).
\eeq
Moreover, according to the discussion above \eqref{eq:abseigen},
$\frac{1}{np_{\max}}\lambda_{\mathsf{K}}(AA^\top)\geq \widetilde c_{\mathsf{K}}^*np_{\max}/\mathsf{K}^2$ holds with probability $1-\mathrm{o}(1)$, where $\widetilde c_{\mathsf{K}}^*>0$ is some constant. This, together with the assumption that $n^{1/3}p_{\max}/\mathsf{K}^2\to\infty$, implies that $\widetilde\lambda_{\mathsf{K}}(A)\gg n^{2/3}$ with probability $1-\mathrm{o}(1)$. Consequently, under the assumption $\lambda_{\mathsf{K}}(A)>0$, with probability $1-\mathrm{o}(1)$,  we have
\[\lambda_{\mathsf{K}}(A)\gg n^{2/3}.\]
This, together with \eqref{eq:eigen_bound} and $\lambda_{\mathsf{K}_0+1}(A)\geq \lambda_{\mathsf{K}}(A)$, as $n\rightarrow\infty$, leads to
\begin{align*}
	\mathbb{P}(\mathbb{T} \leq t_n) 
	&=\mathbb{P}\Bigl(\frac{\lambda_{\mathsf{K}_0+1}(A)-\lambda_{\mathsf{K}_{\max}+1}(A)}{\lambda_{\mathsf{K}_{\max}+1}(A)-\lambda_{\mathsf{K}_{\max}+2}(A)} \leq t_n \Bigr) \\
	&= \mathbb{P}\Bigl[\lambda_{\mathsf{K}_0+1}(A)\leq t_n\bigl\{\lambda_{\mathsf{K}_{\max}+1}(A)-\lambda_{\mathsf{K}_{\max}+2}(A)\bigr\}+\lambda_{\mathsf{K}_{\max}+1}(A) \Bigr] \\
	&= \mathbb{P}\Bigl( \lambda_{\mathsf{K}_0+1}(A)\leq \mathsf{C}_1n^{-1/6+\phi}p^{1/2}_{\max} t_n +\mathsf{C}_2n^{1/2}p_{\max}^{1/2}  \Bigr) \\
	&\le \mathbb{P}\Bigl( \lambda_{\mathsf{K}}(A)\leq  \mathsf{C}_1n^{-1/6+\phi}p^{1/2}_{\max} t_n +\mathsf{C}_2n^{1/2}p_{\max}^{1/2}   \Bigr) \\
	&=\mathrm{o}(1),
\end{align*}
where $\mathsf{C}_1,\mathsf{C}_2>0$ are some constants. The last equality follows from the fact that $\lambda_{\mathsf{K}}(A)\gg n^{2/3}$ with probability $1-\mathrm{o}(1)$, together with the assumption $t_n\ll n^{5/6}$. The latter implies that there exists $\phi>0$ such that  $t_n\leq n^{5/6-\phi}$. Consequently, $ \mathsf{C}_1n^{-1/6+\phi}p^{1/2}_{\max} t_n +\mathsf{C}_2n^{1/2}p_{\max}^{1/2}=\mathrm{O}(n^{2/3})$.
Thus, for a pre-given $\mathsf{K}_{\max}$, we have
\[
\mathbb{P}(\widehat{\mathsf{K}}^* < \mathsf{K}) 
\le \sum_{\mathsf{K}_0=1}^{\mathsf{K}-1} \mathbb{P}(\mathbb{T} \leq t_n) 
\leq \mathsf{K}_{\max}	\mathbb{P}(\mathbb{T} \leq t_n) = \mathrm{o}(1).
\]

\item [(ii)] \textbf{Overestimation.} Under $\mathbf{H}_0: \mathsf{K}=\mathsf{K}_0$, for a pre-given $\mathsf{K}_{\max}$, we have
\begin{align*}
	\mathbb{P}(\mathbb{T} > t_n) 
	&=\mathbb{P}\Bigl(\frac{\lambda_{\mathsf{K}+1}(A)-\lambda_{\mathsf{K}_{\max}+1}(A)}{\lambda_{\mathsf{K}_{\max}+1}(A)-\lambda_{\mathsf{K}_{\max}+2}(A)} > t_n \Bigr) \\
	&= \mathbb{P}\Bigl[\lambda_{\mathsf{K}+1}(A)> t_n\bigl\{\lambda_{\mathsf{K}_{\max}+1}(A)-\lambda_{\mathsf{K}_{\max}+2}(A)\bigr\}+\lambda_{\mathsf{K}_{\max}+1}(A) \Bigr] \\
	&= \mathbb{P}\Bigl( \lambda_{\mathsf{K}+1}(A)> \mathsf{C}_1n^{-1/6+\phi}p^{1/2}_{\max} t_n +\mathsf{C}_2n^{1/2}p_{\max}^{1/2}  \Bigr) \\
	&=\mathrm{o}(1).
\end{align*}
The last equality follows from the fact that, analogous to the derivation of \eqref{eq:eigen_bound}, $\lambda_{\mathsf{K}+1}(A)=\mathrm{O}_{\mathbb{P}}(n^{1/2}p_{\max}^{1/2})$, together with the condition $t_n\asymp n^{\epsilon}$ for some $\epsilon\in(2/3-\phi,5/6)$.
Thus, as $n\rightarrow\infty$, we have
\[ \mathbb{P}(\widehat{\mathsf{K}}^* > \mathsf{K})  \le \mathbb{P}\Big( \mathbb{T} > t_n \Big) = \mathrm{o}(1).\]   
\end{itemize}
The results of (i) and (ii) show that  $\mathbb{P}\left( \hat{\mathsf{K}}^*= \mathsf{K} \right)=1-\mathrm{o}(1)$ as $n\rightarrow\infty$, which completes the proof.

{
\subsection{Supporting lemmas}\label{apen-lemma}
}

\noindent In this section, we present several lemmas, which are utilized in the proofs of the main results.

\bel \label{lemma:invariance_eigen}
For any matrix $B \in \mathbb{R}^{n \times n}$, its singular values remain unchanged under an orthogonal transformation, denoted as $B_*$. Specifically,  the eigenvalues of $B_* B_*^\top$ are identical to those of $BB^\top$.
\eel

\begin{proof} [\bf Proof]
	
	Denote the spectral decomposition of $BB^\top$ as
	\[BB^\top = U \Lambda_B U^\top,\]
	where $U$ is an orthogonal matrix, and $\Lambda_B$ is a diagonal matrix with diagonal entries consisting of the eigenvalues of $BB^\top$. For any orthogonal transformation of $B$, denoted by $B_* $, there exists an orthogonal matrix $S$ such that $B_* = SB$.
	Consequently, we have
	\[B_* B_*^\top = (SB)(SB)^\top = S BB^\top S^\top= S (U \Lambda_B U^\top) S^\top=(SU) \Lambda_B (SU)^\top.\]
	where the matrix $SU$ is also orthogonal. Therefore, the eigenvalues of $B_* B_*^\top$ are identical to those of $BB^\top$, namely $ \Lambda_B$, which completes the proof.
	
\end{proof}

\bel\label{lemma:bound}

Let $D^{(1)}\in\mR^{n\times n}$ be a symmetric non-negative definite matrix, and let $D=\text{diag}(d_1,\cdots,d_{\mathsf{K}},0,\cdots,0)\in\mR^{n\times n}$, where $d_1\ge\cdots\ge d_{\mathsf{K}}>0$.
Denote $M_0$ as the invariant subspace of $D$ associated with the eigenvalue 0, and let $\widetilde M_0$ be the orthogonal projection onto $M_0$. 
Let $\lambda_s(\widetilde M_0D^{(1)}\widetilde M_0|M_0)$ denote the $s$-th largest eigenvalue of $\widetilde M_0D^{(1)}\widetilde M_0$ restricted to the subspace $M_0$.
Define $D(\eta)= D+\eta D^{(1)}$ and  $\widetilde d=d_{\mathsf{K}}/2$, where $\eta$ is a real number satisfying $0<\eta< \widetilde d/\|D^{(1)}\|$. 
Then, it holds that
\[\big|\lambda_{\mathsf{K}+1}(D(\eta))-\eta\lambda_1(\widetilde M_0D^{(1)}\widetilde M_0|M_0)\big|\le 3\widetilde d\frac{|\eta|^2\|D^{(1)}\|^2}{(\widetilde d-|\eta|\|D^{(1)}\|)^2}. \;\; \]
\eel

\begin{proof}  [\bf Proof]
	See  Lemma 1 of \cite{onatski2009formal}.
\end{proof}

\bel \label{lemma:sigular_bound}
Let $A=(A_{ij})_{n\times n}$ denote the adjacency matrix of an undirected network without self-loops, and let $P=(P_{ij})_{n\times n}$ denote the edge probability matrix. For $1\le i<j\le n$, the entries $A_{ij}$ are independently distributed as  $A_{ij}\sim \text{Bernoulli} (P_{ij})$. Assume that  $n \max_{i,j} P_{ij}  \geq c_0 \log n $ for some constant $c_0 > 0$. Then, there exists a constant $\mathsf{C}>0$ such that with probability $1 - \mathrm{o}(1)$
\[ \|A - \mathbb{E}(A)\| \leq \mathsf{C} \sqrt{n \max_{i,j} P_{ij}}. \]

\eel

\begin{proof}  [\bf Proof]
	See  Theorem 5.2 of \cite{lei2015consistency} and Theorem 3.2 of \cite{benaych2020spectral}.
\end{proof}

\bel\label{lemma:hadamard_eigen}
Let $\circ$ denote the Hadamard (entrywise) product, and let $\lambda_s(\cdot)$denote the $s$-th largest eigenvalue of a matrix. For positive semidefinite matrices  $M, B \in\mR^{n\times n}$, we have
\[\lambda_{1}(M\circ B)\leq \max_iM_{ii}\lambda_1(B), \;\; \text{and} \ \  \lambda_{n}(M\circ B)\geq \min_iM_{ii}\lambda_n(B).\]
\eel

\begin{proof}  [\bf Proof]
	See Theorem 5.3.4 on page 316 of \cite{horn1994topics}.
\end{proof}

\bel\label{lemma:hwang}
Let $A=(A_{ij} )_{n\times n}$ be a symmetric adjacency matrix with $\mathsf{K}$ balanced communities. The diagonal entries satisfy $A_{ii}=0$, and the off-diagonal entries $\{A_{ij}: i<j\}$ are independent. Let $q$ denote the square root of the expected average degree. Assume that $\mathbb{E}(A_{ij})=0$, $\sum_{i}\mathbb{E}|A_{ij}|^2=1$, and  $\mathbb{E}|A_{ij}|^m\leq\frac{(cm)^{cm}}{nq^{m-2}}$ for $m\geq 2$ and some constant $c>0$.  Denote by  $\lambda_s(\cdot)$  the $s$-th largest eigenvalue of a matrix.
\begin{itemize}
	{\item [(i). ]  If $1 \ll q \leq n^{1/2}$, for all small $\phi>0$, large $\mathsf{C}>0$ and fixed $i$, as $n\rightarrow \infty$, we have
		\[\mathbb{P}\bigl\{ \big| \lambda_i(A)-\mathsf{L}\big|>n^\phi(q^{-4}+n^{-2/3}) \bigr\} \leq n^{\mathsf{-C}},\]
		where $\mathsf{L}=2+n\kappa_4+\mathrm{O}(q^{-4})$, and $\kappa_4$ denotes the $4$-th cumulant of $A_{ij}$ satisfying $|\kappa_4|=\mathrm{O}(n^{-1}q^{-2})$. }
	
	\item [(ii). ] Furthermore, if $n^{1/6} \ll q \leq n^{1/2}$, for any $x_1\in\mR$, we have
	\[\lim_{n \rightarrow \infty}\mathbb{P}\bigl(n^{2/3}(\lambda_{1}(A)-\mathsf{L})\leq x_1\bigr)=\lim_{n \rightarrow \infty}\mathbb{P}\bigl(n^{2/3}(\lambda_{1}(W)-2)\leq x_1\bigr),\] 
	where $W\in\mR^{n\times n}$ is a GOE matrix.

\end{itemize}

\eel

{
	\begin{proof}  [\bf Proof]
	Part (i) can be obtained from Theorem 2.12 of \cite{hwang2020local} and  Lemma~5.2 of \cite{lee2018local}. Part (ii) can be found in Theorem 2.13 of \cite{hwang2020local}.
	\end{proof}
}

\bel\label{lemma:joint_hwang} 
Under the same conditions as in result (ii) of Lemma~\ref{lemma:hwang}, for any finite $s>0$ and any $x_s\in\mR$, 
\begin{align*}
	&\lim_{n \rightarrow \infty}\mathbb{P}\bigl(n^{2/3}(\lambda_{1}(A)-\mathsf{L})\leq x_1,\cdots, n^{2/3}(\lambda_{s}(A)-\mathsf{L})\leq x_s\bigr)\\
	=&\lim_{n \rightarrow \infty}\mathbb{P}\bigl(n^{2/3}(\lambda_{1}(W)-2)\leq x_1, \cdots, n^{2/3}(\lambda_{s}(W)-2)\leq x_s\bigr).
\end{align*}
\eel

\begin{proof}  [\bf Proof]
	See Remark 3.4 of \cite{lee2014necessary}, Remark 2.11 of \cite{lee2018local}, and equation (2.42) of \cite{erdHos2012spectral}.
\end{proof}

\bel \label{lemma:eigsA_cards}
Let $A=(A_{ij} )_{n\times n}$ be a symmetric adjacency matrix with $\mathsf{K}$ balanced communities. The diagonal entries satisfy $A_{ii}=0$, and the off-diagonal entries $\{A_{ij}: i<j\}$ are independently distributed as  $A_{ij}\sim \text{Bernoulli} (P_{ij})$, where $P=(P_{ij})_{n\times n}$ denotes the edge probability matrix. Denote by $\widetilde\lambda_s(\cdot)$  the $s$-th largest eigenvalue in absolute value. Suppose that $\min_{i,j}P_{ij}\gg n^{-1}, \max_{i,j}P_{ij}=\mathrm{O}(1)$ and $\widetilde\lambda_{\mathsf{K}}(P)\gg n^{1/2}$. Then, for some sufficiently small constant $\delta > 0$ and large positive integer $\mathsf{C}$, the following holds with probability $1 - \mathrm{o}(1)$,
\[ \# \bigl\{ i : \widetilde\lambda_i(A) \geq \delta \bigr\} \geq \mathsf{K} + \mathsf{C},\]
where $\#\{\cdot\}$ denotes the cardinality of the set.
\eel

\begin{proof}  [\bf Proof]
The proof proceeds by considering two cases: $i\leq \mathsf{K}$ and $i>\mathsf{K}$. 
For $i \le \mathsf{K}$, we show that, with probability $1 - \mathrm{o}(1)$, $\widetilde\lambda_i(A)$ diverges, and is therefore exceed any positive constant $\delta$.
For $i > \mathsf{K}$, the result (i) of Lemma~\ref{lemma:hwang} implies that, for each fixed $i$, there exists a sufficiently small constant $\delta > 0$ such that $\widetilde\lambda_i(A)\geq \delta$ holds with probability $1 - \mathrm{o}(1)$.
Consequently, there exist constants $\delta$ and $\mathsf{C}$ such that,  with probability $1 - \mathrm{o}(1)$, at least $\mathsf{K} + \mathsf{C}$ eigenvalues of $A$ are bounded away from $\delta$.

\begin{enumerate}
	\item [(i) ]  \textbf{For $i\leq\mathsf{K}$.}	Let $\widetilde{A} = A - \mathbb{E}(A)$ be the centered noise matrix. 
	Applying Lemma~\ref{lemma:sigular_bound}, we can obtain
   \beq\label{eq:lema_opA}
	\| \widetilde{A} \| = \mathrm{O}_{\mathbb{P}}\bigl(n^{1/2} \bigr).
	\eeq
	Given $\mathbb{E}(A)= P-\text{diag}( P)$, we can write $A=P+\widetilde{A}-\text{diag}( P)$.  Denote by  $\lambda_s(\cdot)$  the $s$-th largest eigenvalue of a matrix. Then, by Weyl's inequality, we have
	\[	|\lambda_i(A) - \lambda_i (P)|\leq \|\widetilde{A}-\text{diag}(P)\|. \]
	This, together with the assumption that $\widetilde\lambda_{\mathsf{K}}(P) \gg n^{1/2}$ and \eqref{eq:lema_opA}, we have, with probability $1-\mathrm{o}(1)$,
	\[ \widetilde\lambda_{\mathsf{K}}(A) \gg n^{1/2}. \]
	Consequently, $\widetilde\lambda_i(A) \ge \delta$ holds with probability $1 - \mathrm{o}(1)$ for any constant $\delta > 0$.

 {
	\item [(ii)]  \textbf{For $i>\mathsf{K}$.} 	Recall that $A=P+\widetilde{A}-\text{diag}( P)$. By Weyl’s inequality,
	\[	|\lambda_i(A) - \lambda_i (\widetilde{A}+P)|\leq \|\text{diag}(P)\|=\max_{i}P_{ii}. \]
	Consequently, 
	\[\lambda_i(A) = \lambda_i (\widetilde{A}+P)+\mathrm{O}(\max_{i}P_{ii}). \]
	Applying Weyl’s inequality again yields $ \lambda_i (\widetilde{A}+P)\geq \lambda_i (\widetilde{A})+\lambda_{n}(P)$.
	This leads to
	\[\lambda_i(A) \ge\lambda_i (\widetilde{A})+\lambda_{n}(P)+\mathrm{O}(\max_{i}P_{ii}). \]
	Since the rank of $P$ is $\mathsf{K}$, it follows that  $\lambda_{n}(P)=0$. Therefore, 
	\beq\label{lemma:igeqK}
	\lambda_i(A) \ge\lambda_i (\widetilde{A})+\mathrm{O}(\max_{i}P_{ii}).
	\eeq
Next, we introduce the auxiliary matrix
$\widetilde{A}\circ\widetilde\Omega$, where $\circ$ denotes the Hadamard product and $\widetilde\Omega=(\widetilde\Omega_{ij})_{n\times n}$ is defined as \[\widetilde\Omega_{ij}=\{(n-1)P_{ij}(1-P_{ij})\}^{-1/2}.\] 
Let $q=n^{-1/2}(\sum_{i,j}P_{ij})^{1/2}$. Under the assumption $A_{ij}\sim \text{Bernoulli} (P_{ij})$, one can verify that for $1\le i\not=j\le n$,
\[\mathbb{E}( {\widetilde A}_{ij}\circ\widetilde\Omega_{ij})=0, \; \; \sum_{i}\mathbb{E}| {\widetilde A}_{ij}\circ\widetilde\Omega_{ij}|^2=1,\]
and for all  $m\ge 2$  
\[ \mathbb{E}|{\widetilde A}_{ij}\circ\widetilde\Omega_{ij}|^m\le\frac{(cm)^{cm}}{nq^{m-2}},\]
where $c>0$ is some constant.
Under these  moment conditions, together with the assumptions that $\min_{i,j}P_{ij}\gg n^{-1}$ and that the network is partitioned into balanced communities,  the result (i) of  Lemma~\ref{lemma:hwang} implies that, for fixed $i$, $\lambda_{i}(\widetilde{A}\circ\widetilde\Omega)\to\mathsf{L}$ with probability $1-\mathrm{o}(1)$, where $\mathsf{L}-2=\mathrm{o}(1)$.
Since $\widetilde\Omega_{ij}\to 0$ uniformly in $1\le i, j \le n$,  $\widetilde{A}\circ\widetilde\Omega$ is a vanishing rescaling of $\widetilde{A}$. Therefore,  $\lambda_{i}(\widetilde{A})$ is asymptotically larger in magnitude than $\lambda_{i}(\widetilde{A}\circ\widetilde\Omega)$. Consequently, for fixed $i$, there exists some constant $0<\delta<\mathsf{L}$ such that $\lambda_i(\widetilde A) \geq \delta$ with probability $1-\mathrm{o}(1)$.  By \eqref{lemma:igeqK}, the $\delta$ can be chosen such that $\lambda_i(A) \geq \delta$ with probability $1-\mathrm{o}(1)$.
}
\end{enumerate}

Combining the discussions of cases (i) and (ii), we conclude that, with probability  $1 - \mathrm{o}(1)$, there exists a constant $\delta > 0$ such that $\widetilde\lambda_i(A) \geq \delta$ for $i=1,\cdots,\mathsf{K} + \mathsf{C}$, where $\mathsf{C}>0$ is a fixed integer. This completes the proof.

\end{proof}

\section{Algorithm for determining an $\mathsf{K}_{\max}$}

\noindent
In this section, we present the detailed procedure for computing $\mathsf{K}_{\mathrm{PA}}$ in Algorithm~\ref{alg:PA}, following \cite{Dobriban2020}. Based on this, in (\ref{test-T}) we set $\mathsf{K}_{\max} = \mathsf{K}_{\mathrm{PA}} + 5$.

\begin{algorithm}[!ht]
	\caption{Parallel analysis for selecting $\mathsf{K}_{\text{PA}}$}
	\label{alg:PA}
	
	\KwIn{Adjacency matrix $A \in \mathbb{R}^{n\times n}$, number of permutations $\mathsf{B}$,  percentile level $q$.}
	
	\noindent {\bf 1. Compute the eigenvalues of $A$:} $\lambda_1(A) \ge \lambda_2(A) \ge \cdots \ge \lambda_n(A)$.
	
	\noindent {\bf 2. Permutation-based reference distribution.}
	\begin{enumerate}
		\item[(i).] \For{$b = 1$ \KwTo $\mathsf{B}$}{
			\begin{enumerate}
				\item[(a).] Generate a permutation array
				\[
				\bm\tau^{(b)} = (\tau^{(b)}_1, \tau^{(b)}_2, \ldots, \tau^{(b)}_n),
				\]
				where each $\tau^{(b)}_j$ is an independent random permutation of $\{1,\ldots,n\}$ used to permute the $j$-th column of $A$. Define the permuted matrix $A_{\bm\tau}^{(b)}$ with entries
				\[
				(A_{\bm\tau}^{(b)})_{ij} = A_{\tau^{(b)}_j(i), j}, \qquad 1 \le i,j \le n.
				\]
				\item[(b).] Compute the eigenvalues of $A_{\bm\tau}^{(b)}$: $\lambda_1(A_{\bm\tau}^{(b)}) \ge \cdots \ge \lambda_n(A_{\bm\tau}^{(b)})$.
				
			\end{enumerate}
		}
		
		\item[(ii).] For each $j = 1, \cdots, n$, let $q_j$ be the empirical $q$-quantile of 
		$\{\lambda_j(A_{\bm\tau}^{(b)}) : b = 1,\ldots,\mathsf{B}\}$.
		
	\end{enumerate}
	
	\KwOut{	$\mathsf{K}_{\mathrm{PA}} = \max \bigl\{ j \in \{1, \cdots, n\} : \lambda_j(A) > q_j \bigr\}$.}
	
\end{algorithm}

We remark that, for convenience, in Algorithm \ref{alg:PA}, we choose the percentile level $q$ and the number of permutations $\mathsf{B}$ such that $q\mathsf{B}$ is an integer. It is also worth noting that, in the practical implementation of Algorithm~\ref{alg:PA} for large-scale networks, the computation can be accelerated by selecting $\mathsf{K}_{\mathrm{PA}}$ from among the first $\mathsf{C}$ largest eigenvalues of $A$, rather than computing all $n$ eigenvalues, where $\mathsf{C}$ is a sufficiently large constant, for example, $\mathsf{C}=\lfloor \sqrt{n} \rfloor$.

\section{ Additional numerical results}\label{apen-sec5}

\subsection{Analysis of computational efficiency}\label{sec_computationaleffiency}

\begin{figure}[htbp]
\centering
\includegraphics[scale=1.0]{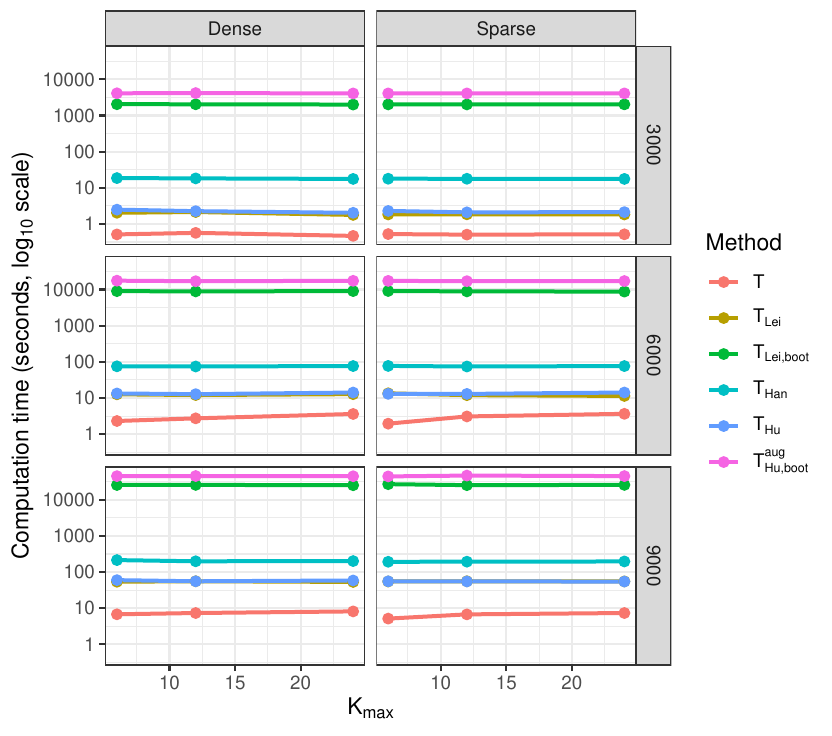} 
\caption{Computation time (in seconds) for dense and sparse SBMs. Networks have equally sized communities with $n \in \{3,000, 6,000, 9,000\}$, $\mathsf{K}_0 =5$, and $\mathsf{C}=2, 5, 10$.}
\label{fig:cpu}
\end{figure}

{
\noindent
This section reports the computation time of $\mathbb{T}$ along with  $\mathbb{T}_{\text{Lei}}$, $\mathbb{T}_{\text{Lei,boot}}$, $\mathbb{T}_{\text{Han}}$, $\mathbb{T}_\text{Hu}$ and $\mathbb{T}_{\text{Hu,boot}}^{\text{aug}}$. 
It is worth noting that the complexity of finding all $n$ eigenvalues of $A$ is roughly ${\mathrm{O}}(n^3)$, and this can be computationally costly. In contrast, our test statistic ${\mathbb{T}}$, for any given ${\mathsf{K}_{\max}}$, only requires calculating the first ${\mathsf{K}_{\max}}+2$ eigenvalues of $A$. Hence, the computational complexity can be reduced to $\mathrm{O}(n^2\mathsf{K}_{\max})$ or even to $\mathsf{O}(\mathsf{K}_{\max})$ if the network is sparse. Therefore, for $\mathbb{T}$, the computational cost primarily depends on the network size, sparsity, and the value of $\mathsf{K}_{\max}$, but remains independent of the specific model structure. 

We conduct simulations for dense and sparse SBMs under the same settings as those used in Section \ref{sec-simu}, except that we consider network sizes $n = 3,000, 6,000$, and $9,000$. We fix $\mathsf{K}_0 = 5$, and  choose $\mathsf{K}_{\max}=\mathsf{K}_{\text{PA}}+\mathsf{C}$ with $\mathsf{C}=2, 5, 10$. The computation time (in seconds) is obtained on a MacBook with an M3 Pro chip and 18~GB of RAM. 
We note that the critical values of our test statistic $\mathbb{T}$ are precomputed separately for various combinations of $\mathsf{K}_0$, $\mathsf{K}_{\max}$, and $n$. Consequently, in our numerical analyses, $\mathbb{T}$ is directly compared with the corresponding precomputed critical value, and the reported computation time excludes the cost of computing the critical values.

Based on the results in Figure~\ref{fig:cpu}, we observe that the computation cost is primarily determined by the network size $n$, and shows only mild sensitivity to increases in $\mathsf{K}_{\max}$ and network density. Across all configurations, our proposed method $\mathbb{T}$ consistently exhibits the lowest computation time, whereas the competing methods incur substantially higher computational costs. In particular, $\mathbb{T}_{\text{Lei, boot}}$ and $\mathbb{T}^{\text{aug}}_{\text{Hu,boot}}$ are several orders of magnitude slower, with computation times reaching tens of thousands of seconds. In contrast, the runtime of {$\mathbb{T}$} remains at the sub-second to single-digit--second level even when $n$ increases to $9,000$ under both dense and sparse networks. Therefore, our proposed method demonstrates greater robustness and practical scalability for large-scale network analysis.

}

\subsection{Simmons College network data analysis}\label{sec_simmons}

\noindent
In this section, we consider an additional real network that contains all friendship links among Facebook users at Simmons College, recorded on a specific day in September 2005 \citep{traud2012social}. It consists of 1,137 nodes, with an edge density of approximately 3.76\%.
{\citet{traud2012social} observed that the community structure of this network is strongly correlated with graduation year, meaning that students graduating in the same year are more likely to be friends. Hence, the true number of clusters in this network can be considered as $\mathsf{K}=2$.} It was pointed out in \cite{jin2022improvements} that the community structure in the Simmons College network is very weak and community detection algorithms  usually yield high misclassification errors.

{We employ our test statistic $\mathbb{T}$, along with $\mathbb{T}_{\text{Lei, boot}}$, $\mathbb{T}_{\text{Hu, boot}}^{\text{aug}}$ and $\mathbb{T}_{\text{Han}}$, to test $\mathsf{K}_0 = 1, 2$, as all tests with $\mathsf{K}_0>2$ yield the same conclusion. The testing results, summarized in Table~\ref{simmons}, show that both $\mathbb{T}$ and $\mathbb{T}_{\text{Han}}$ reject $\mathbf{H}_0: \mathsf{K} = 1$, but do not reject $\mathbf{H}_0: \mathsf{K} = 2$, identifying at least two communities. In contrast, $\mathbb{T}_{\text{Lei, boot}}$ and $\mathbb{T}_{\text{Hu, boot}}^{\text{aug}}$ reject all null hypotheses. Therefore, although the community structure in the Simmons College network is weak, our method remains applicable and effective for identifying $\mathsf{K}$. }

\begin{table}[h!]
	\centering
	\renewcommand\arraystretch{1.5}{
		\begin{tabular}{ccccc}
			\hline
			\textbf{$\mathsf{K}_0$} & \textbf{$\mathbb{T}$} & \textbf{$\mathbb{T}_{\text{Han}}$} & \textbf{$\mathbb{T}_{\text{Lei, boot}}$} & \textbf{$\mathbb{T}_{\text{Hu, boot}}^{\text{aug}}$} \\ \hline
			\textbf{1} & Reject & Reject & Reject & Reject \\
			\textbf{2} & \bf{Accept} & \bf{Accept} & Reject & Reject \\ 
			\hline
	\end{tabular}}
		\caption{Testing results for Simmons College network.}
	\label{simmons}
\end{table}

%
%
%
%

\section{Some additional remarks}

%
%
%
%
%

\begin{remark}\label{sec_techinicaldiscussion}
It is worth noting that two key results used in proving Theorem \ref{theory-null} in Section  \ref{apen-sec3} are Lemma 1 of \cite{onatski2009formal} and Theorem 2.13 of \cite{hwang2020local}.
Specifically, we first apply Lemma 1 of \cite{onatski2009formal}  to carefully bound the differences	between $c_{i^*j^*}\tilde{\Omega}_{G_I,i^* j^*}\lambda_k(A)$ and {$\lambda_{k-\mathsf{K}}(\tilde A_{G_I}\circ\tilde{\Omega}_{G_I})$} for any $k\ge {\mathsf{K}}+1$, as stated in \eqref{eq:AR22-star}. Here, $c_{i^*j^*}$ and $\tilde{\Omega}_{G_I,i^* j^*}$ are scaling factors, and $\tilde A_{G_I}\circ\tilde{\Omega}_{G_I}$ represents the variance-normalized version of ${\tilde A}_{G_I}$, which is an $n-{\mathsf{K}}_0$ principal submatrix of the noise matrix $\tilde{A}$.
However, \cite{onatski2009formal} assumes that ${\mathsf{K}}$ is finite and that both $A$ and $\tilde A_{G_I}$ are dense and continuous matrices. This presents challenges in extending {Lemma 1 of \cite{onatski2009formal}} to the binary sparse matrices $A$ and $\tilde A_{G_I}$ considered in this paper, especially when ${\mathsf{K}}$  is divergent. 
Then, we utilize Theorem 2.13 of \cite{hwang2020local} to demonstrate that the distribution of {$n^{2/3}(\lambda_{k-\mathsf{K}}(\tilde A_{G_I}\circ\tilde{\Omega}_{G_I})-\mathsf{L})$} converges to the type-I Tracy-Widom distribution via Airy kernel by establishing higher-order moment conditions for {$\tilde A_{G_I}\circ\tilde{\Omega}_{G_I}$,} where {$\mathsf{L}$} is a location parameter that is independent of $k$. This, in turn, imposes higher-order moment requirements on the construction of $\tilde A_{G_I}$ in the previous step.
\end{remark}



\begin{remark}
	
Although existing studies \citep{lei2016goodness, hu2021using, han2023universal} also involve a trade-off between network sparsity and the divergence rate of $\mathsf{K}$, our work provides an explicit, theoretically justified criterion based on $n^{1/3}\max_{i, j}P_{ij} / \mathsf{K}^2 \to \infty$. This criterion surpasses existing methods by allowing either greater network sparsity or a higher admissible growth rate of $\mathsf{K}$.
It is worth noting that this criterion is necessary to ensure that the distribution of $n^{2/3}\big(\lambda_{k-\mathsf{K}}(\tilde A_{G_I}\circ\tilde{\Omega}_{G_I}) - \mathsf{L}\big)$
converges to the type-I Tracy–Widom distribution via Airy kernel, which underlies the implementation of the hypothesis test \eqref{test-hypo}. An illustration is provided in Figure \ref{fig:tradeoff} below.

\begin{figure}[htbp]
	\centering
	\includegraphics[scale=0.55]{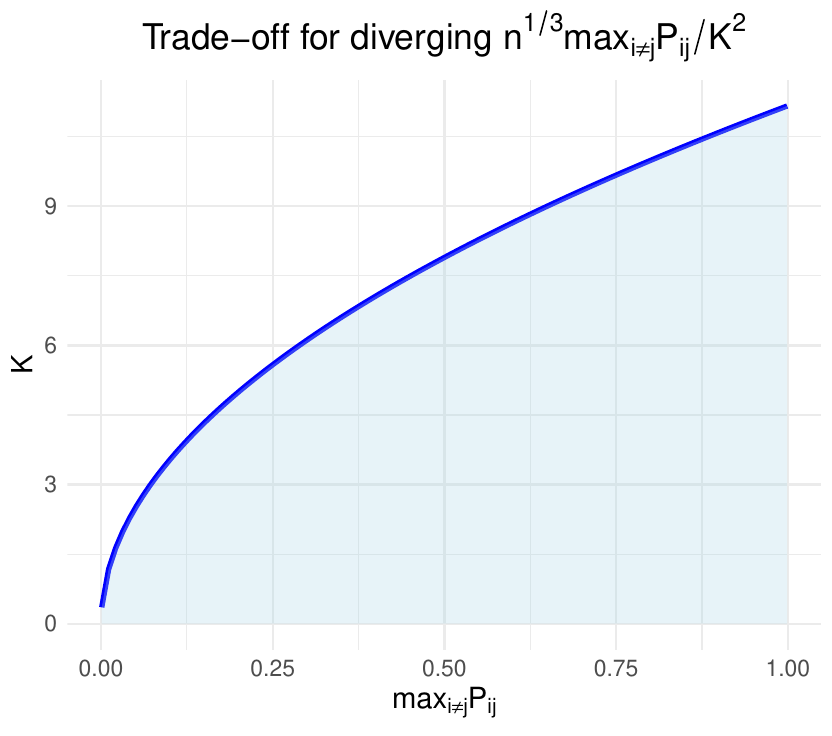} 
	\caption{Feasible region for {$n^{1/3}\max_{i, j}P_{ij} / \mathsf{K}^2 =3n^{\epsilon_1}$}, where $n=3,000$ and  $\epsilon_1=0.005$. The shaded blue area represents the feasible values of ${\mathsf{K}}$ corresponding to each {$\max_{i, j}P_{ij}$,}  while the solid blue line shows the maximal value that ${\mathsf{K}}$ can take.}
	\label{fig:tradeoff}
\end{figure}
\end{remark}

\begin{remark}\label{rmk_techinicalremark}
Two technical remarks are in order. First, 	
the assumption  $\min_{1\le k\le \mathsf{K}} |\lambda_k(P)|\ge cn\max_{i , j}P_{ij}/\mathsf{K}$ arises from two core requirements underlying Theorem~\ref{theory-null}: (i) the $\mathsf{K}$ leading eigenvalues of $A$ are dominated by $P$, while the remaining eigenvalues follow those of the centered noise matrix $\widetilde{A}$;  (ii) the remaining $\mathsf{K}_{\max} - \mathsf{K} + 2$ eigenvalues  of $A$ converge in distribution to the Tracy–Widom law via the Airy kernel. Requirement (i) necessitates that $\min_{1\le k\le \mathsf{K}} |\lambda_k(P)|$ exceeds the operator norm of $\widetilde{A}$.  Requirement (ii) further requires that the convergence rate of these $\mathsf{K}_{\max} - \mathsf{K} + 2$ eigenvalues to the centered noise spectrum, which is shown to be governed by $\min_{1\le k\le \mathsf{K}} |\lambda_k(P)|$,  be faster than the fluctuation scale of the Tracy–Widom distribution, i.e., $\mathrm{O}(n^{-2/3})$.

Second,	to invoke Theorem 2.13 of \cite{hwang2020local}  and establish that $\lambda_{k-\mathsf{K}}(\tilde A_{G_I}\circ\tilde{\Omega}_{G_I})$ is asymptotically Tracy–Widom distributed via Airy kernel, thereby enabling the use of the Tracy–Widom law to characterize the limiting distribution of $\mathbb{T}$, we impose a mild requirement that the network is relatively balanced across communities. Specifically, for $1\le u\le \mathsf{K}$, let $n_k=\sum_{i=1}^n\pi_{i,k}$ denote the number of nodes (possibly) located in community $k$.
	We require $c^*_1\le\min_{k,l}n_k/n_l\le\max_{k,l}n_k/n_l\le c^*_2$ for some positive constants $c^*_1$ and $c^*_2$. 
\end{remark}


%
%
%
%
%
%

\bibliographystyle{chicago}

\bibliography{paper-ref}

\end{document}